\documentclass[lettersize,journal]{IEEEtran}
\usepackage{amsmath,amsfonts}
\usepackage{array}
\usepackage[caption=false,font=normalsize,labelfont=sf,textfont=sf]{subfig}
\usepackage{textcomp}
\usepackage{stfloats}
\usepackage{url}
\usepackage{verbatim}
\usepackage{graphicx}
\usepackage{cite}

\usepackage{algorithm}
\usepackage{algpseudocode}
\usepackage{cite}
\usepackage{amsmath,amssymb,amsfonts}
\usepackage{graphicx}
\usepackage{textcomp}
\usepackage{xcolor}
\usepackage{colortbl}
\usepackage{multicol,cite,url,color}
\usepackage{comment}
\usepackage{tabularray}
\usepackage{tabularx}
\usepackage{lettrine}
\definecolor{LightGreen}{rgb}{0.75, 1.0, 0.75} 
\definecolor{LightRed}{rgb}{1.0, 0.8, 0.8}
\begin{document}

\title{Radiance Field Delta Video Compression \\in Edge-Enabled Vehicular Metaverse}

\author{\IEEEauthorblockN{Mat\'{u}\v{s} Dopiriak\textsuperscript{1}, Eugen \v{S}lapak\textsuperscript{1}, Juraj Gazda\textsuperscript{1}, Devendra Singh Gurjar\textsuperscript{2},\\Mohammad Abdullah Al Faruque\textsuperscript{3}, Marco Levorato\textsuperscript{3}} \\\IEEEauthorblockA{\textsuperscript{1}Department of Computers and Informatics, Technical University of Ko\v{s}ice, Slovakia\\ Email: \{matus.dopiriak, eugen.slapak, juraj.gazda\}@tuke.sk} 
\\\IEEEauthorblockA{\textsuperscript{2}Department of Electronics and Communications Engineering, National Institute of Technology Silchar, India}
Email: dsgurjar@ece.nits.ac.in
\\\IEEEauthorblockA{\textsuperscript{3}Department of Electrical Engineering and Computer Science, University of California, Irvine, United States}
Email: \{alfaruqu, levorato\}@uci.edu}



\maketitle

\begin{abstract}
Connected and autonomous vehicles (CAVs) offload computationally intensive tasks to multi-access edge computing (MEC) servers via vehicle-to-infrastructure (V2I) communication, enabling applications within the vehicular metaverse, which transforms physical environment into the digital space enabling advanced analysis or predictive modeling. A core challenge is physical-to-virtual (P2V) synchronization through digital twins (DTs), reliant on MEC networks and ultra-reliable low-latency communication (URLLC). To address this, we introduce radiance field (RF) delta video compression (RFDVC), which uses RF-encoder and RF-decoder architecture using distributed RFs as DTs storing photorealistic 3D urban scenes in compressed form. This method extracts differences between CAV-frame capturing actual traffic and RF-frame capturing empty scene from the same camera pose in batches encoded and transmitted over the MEC network. Experiments show data savings up to 71\% against H.264 codec and 44\% against H.265 codec under different conditions as lighting changes, and rain. RFDVC also demonstrates resilience to transmission errors, significantly outperforming the standard codec in non-rainy conditions with up to a +0.26 structural similarity index measure (SSIM) improvement over H.264 codec, and maintaining a +0.18 SSIM improvement even in challenging rainy conditions, both measured at a block error rate (BLER) of 0.25.
\end{abstract}


\section{Introduction}
\IEEEPARstart{C}{onnected} and autonomous vehicles (CAVs) are at the forefront of modern transportation systems, relying on advanced perception module to process terabytes of data from sensors (e.g., LiDARs or cameras). The module interprets the environment to ensure safe and efficient navigation by executing computationally intensive tasks such as localization, object detection, scene understanding, prediction, and tracking in real-time \cite{longmilestones2023}. To manage these high demands, CAVs employ vehicle-to-infrastructure (V2I) communication, offloading processing tasks to edge computing servers \cite{islamsurvey2021}.\par

Tech giants have recently invested heavily towards transforming physical environment into the digital space within the metaverse \cite{khan2024metaverse}, including applications in transportation systems, known as vehicular metaverse. It integrates real-time traffic information into a unified virtual space, enabling advanced analysis, predictive modeling, and efficient resource management. A key challenge in this domain is the physical-to-virtual (P2V) synchronization \cite{xuepvisa2024}, which relies on V2I communication utilizing multi-access edge computing (MEC) networks \cite{yuspotlighter2024} and ultra-reliable low latency communications (URLLC) \cite{alves2022beyond} requiring reliability levels up to $1 - 10^{-5} \%$ and end-to-end latency in the range of $5-10$ ms.
In recent years, deep learning (DL) techniques have been employed to enhance P2V synchronization and meet stringent communication requirements through video compression \cite{ding2021advances, birman2020overview}. These techniques are primarily used in conjunction with standard video codecs for intra-prediction, focusing on compressing intra frames (complete frames), which contribute the most to the overall bitrate. Additionally, other approaches use implicit neural representations to transform individual frames \cite{zhang2021implicit, hao2021nerv, ghorbel2024ner++} or use learning-based compression of both temporal and binocular redundancy in stereo video \cite{chen2022lsvc}.

The metaverse endeavors aim at closing this synchronization gap by developing photorealistic digital environments (i.e., digital twins (DTs)) at the edge of the MEC network \cite{wudigital2021, hashash2022towards, khan2022digital}. Recent studies as in \cite{mihai2022digital, minrui2023full, ren2022quantum, xu2023generative, zhou2023vetaverse} also underscore the critical role of DTs at the edge in reconstructing alternative digital environments that do not require advanced sensor data, such as LiDARs, nor their redundant transmission over the network.\par

Since 2020, neural radiance fields (NeRFs) \cite{mildenhall2020nerf} have become a significant breakthrough in a 3D scene reconstruction and rendering, enabling potential DT applications without relying on LiDAR data transmission over the MEC network. This is achieved through implicit depth prediction via volume rendering along a ray and the development of advanced NeRF variants. NeRFs enable the generation of novel views from previously unseen angles and provide efficient compression of 3D scenes, even when trained on sparse input data with varying camera poses. The underlying volumetric representation is encoded as a multi-layer perceptron (MLP) and rendered through volumetric ray-marching. NeRFs excel in modeling volumetric radiance, which is crucial for accurately capturing intricate details, subtle lighting effects, and complex lighting conditions, such as advanced reflections and dynamic weather changes, essential for a realistic scene reconstruction. In recent years, significant advancements have been made in reducing training and rendering time while improving the accuracy of radiance fields (RFs) as in \cite{barron2021mipnerf, barron2023zipnerf}, especially instant neural graphics primitives (INGP) \cite{mullerinstant2022} and 3D gaussian splatting (3DGS) \cite{kerbl20233d} explained more in-depth in Section \ref{sub:rfs_ddt}.\par

In the survey on the application of RFs in autonomous driving (AD) \cite{heneural2024}, traditional RF methods are highlighted as struggling with scalability, leading to visual artifacts and reduced fidelity in large outdoor environments. To address this issue, Block-NeRF \cite{tancikblock2022} was proposed, decomposing extensive scenes into multiple compact, independently trained NeRFs. 
While recent advancements offer promising solutions to large-scale scenes, they are hindered by significant computational demands. Streamable memory efficient RF (SMERF) \cite{duckworth2024smerf} introduces a hierarchical model partitioning scheme and distillation training, while NeRF-XL \cite{li2024nerfxl} distributes tasks across multiple GPUs. Hierarchical 3DGS \cite{hierarchicalgaussians24} partitions large scenes into chunks, optimizing each independently. \par

RF-based simulators \cite{pun2023neural, zhou2024drivinggaussian, wu2023mars} are utilizing scene decomposition of dynamic elements and scene relighting to be primarily used for further simulations in AD or secondarily be used as DTs. However, the experimental results lack robustness for generalization across diverse scenarios and depend on LiDAR data \cite{chunmao2024largescale, xiuzhong2024pcnerf} for computationally intensive training. Distributed visual data collection for creating NeRF-based DT for autonomous mobility was examined in \cite{liu2023visualization}. The work has tested the speed of real-world to DT updates when considering different DT quality and network conditions. While this work tests many of the assumptions crucial for our work, it misses the key idea of use of NeRFs for rapid compression to mitigate the latency. \par 

In this study, we extend our previous work in \cite{slapakdistributed2024} by introducing RF delta video compression (RFDVC). This approach leverages a V2I communication between a CAV and a MEC server. RFs, distributed across the CAV (sender) and the MEC server (receiver), reconstruct static elements of an empty urban scene (e.g., buildings, roads or traffic lights), reducing the need for their repeated transmission.
We segment differences, between CAV-frame capturing actual traffic and RF-frame capturing an empty scene from the same pose, represented as Delta-frames. The sender encodes and transmits the Delta-frames in batches over the MEC network, while the receiver reconstructs the CAV-frame as a Rec-frame by applying the Delta-frame to the corresponding RF-frame.
Standard codecs face challenges encoding complex patterns within or across frames, leading to increased bitrate and potential MEC network congestion. RFDVC minimizes redundant data by storing it in distributed RFs at both ends, while Delta-frames retain only differences between CAV-frame and RF-frame from the same camera pose. The remaining pixels are masked out as a solid black background enabling more efficient compression. RFDVC achieves notable throughput savings and reduces latency, which is pivotal for a real-time communication. \par

A critical distinction of our approach lies in its evaluation across a broad spectrum of environmental and operational conditions. Unlike prior works that often focus on static or idealized scenarios, we systematically test performance under varying weather conditions (e.g., rain, wet surfaces), different times of day (morning, noon, evening), and dynamic traffic scenarios, such as sparse and dense traffic. This comprehensive evaluation demonstrates the robustness and adaptability of RFDVC, addressing gaps in previous research that often overlook these real-world complexities. Our findings highlight the capability of RFDVC to maintain efficient communication and accurate results even in challenging and diverse conditions, setting it apart from existing compression frameworks.

The main contributions
of this paper are summarized as follows:

\begin{enumerate}
    \item  We present an RF-encoder and an RF-decoder architecture using distributed RFs as DTs for storing photorealistic 3D urban scenes in a compressed form.

    \item We propose delta segmentation (DS) algorithm for extraction of differences between CAV-frame and RF-frame pairs without being limited by a predefined set of classes using zero-shot generalization. We also guarantee the transmission of critical classes (e.g., pedestrians or vehicles) using semantic segmentation.
    

    \item We validate the data compression achieved by the RFDVC framework compared to H.264 and H.265 codecs under different conditions as lighting changes, and rain. We also evaluate its performance under simulated packet loss to assess quality under challenging network conditions.
\end{enumerate}

\section{Video Compression Preliminaries}
In this section, we present the foundational principles of the H.264 and H.265 codecs widely utilized for real-time V2I communication to achieve efficient video compression via inter-frame and intra-frame encoding techniques. Subsequently, we introduce RFs as compressed representations of 3D scenes.

\subsection{Standard Video Compression}
Standard video codecs like H.264 and H.265 (HEVC) are critical for real-time V2I communication. These codecs use spatial compression to eliminate redundancy within individual frames and temporal compression to minimize redundancy across multiple frames. Frames are generally organized into a group of pictures (GOP) structure, defined as the interval between two complete frames, known as intra-frames. The GOP consists of both intra-frames (I-frames) and inter-frames. I-frames store the full image information, similar to a standard jpeg file. Inter-frames, which depend on I-frames for reconstruction, are categorized into P-frames and B-frames. P-frames store differences from preceding frames, while B-frames store differences from both preceding and subsequent frames. B-frames achieve the highest compression ratio and have the smallest file size on average, among these frame types. Fig.~\ref{fig:cav_mec}~b) shows a sequence of encoded frames transmitted over the network within a GOP structure. It highlights the arrangement of B-frames and P-frames between two I-frames. \par

Traditional video compression depends on frequent I-frame transmission to maintain video quality, especially in scenes with high motion or complexity, as P-frames and B-frames rely on I-frames for reconstruction. Extending the GOP length can reduce the overall data size by decreasing the number of I-frames, which are the largest contributors to the GOP size. However, this approach may reduce frame reconstruction quality and increase vulnerability to errors caused by network issues, such as packet loss or bit flips. To balance compression efficiency, video quality, and latency, careful management of I-frame transmission frequency, or even avoiding frequent I-frame transmission, is necessary to prevent network congestion. In the context of CAVs, maintaining low latency remains a crucial challenge, and the compression of I-frames is still a significant bottleneck in standard video coding (VC) when deployed for real-time data transmission.

\subsection{Radiance Fields as Compressed 3D Scenes}
\label{sub:rfs_ddt}
We explore the core principles of RFs pivotal to our approach for efficient and photorealistic compression of 3D scenes used as DTs in the vehicular metaverse. We focus on two RF variants used in our experiments: INGP \cite{mullerinstant2022} and 3DGS \cite{kerbl20233d}. These variants were selected primarily for their favorable training time and reasonable accuracy they offer in reconstructing 3D scenes. \par

NeRFs \cite{mildenhall2020nerf} utilize MLPs to encode 3D scenes, parametrizing images with camera poses and optimizing a volumetric scene function approximated by MLP \( \Phi_{\Theta} \):
\begin{equation}
\Phi_{\Theta} :(\boldsymbol{x}, \boldsymbol{d}) \rightarrow (\boldsymbol{\hat{c}}, \sigma),
\end{equation}

\noindent where \( \boldsymbol{x} \) is the location of the point in 3D space, \( \boldsymbol{d} \) is the viewing direction, \( \boldsymbol{\hat{c}} \) is the emitted radiance, and \( \sigma \) is the volume density.

Rendering from RF involves calculating the expected color \( \hat{C}(\boldsymbol{r}) \) of a ray \( \boldsymbol{r}(t) = \boldsymbol{o} + t\boldsymbol{d} \), defined by its origin $\boldsymbol{o}$ and direction $\boldsymbol{d}$, within near and far bounds $t_n$ and $t_f$, by integrating the transmittance \( T(t) \): 

\begin{equation}
\hat{C}(\boldsymbol{r}) =\int_{t_n}^{t_f}T(t)\sigma(\boldsymbol{r}(t))\boldsymbol{\hat{c}}(\boldsymbol{r}(t),\boldsymbol{d})dt.
\end{equation}

The transmittance $T(t)$ is defined as follows:
\begin{equation} \label{eq:transmittance}
T(t) = exp(-\int_{t_n}^{t}\sigma(\boldsymbol{r}(s))ds),
\end{equation}
\noindent where $T(t)$ is the probability that ray in the range from  $t_n$ to $t$ does not hit any material.

The loss \( \mathcal{L}_{NeRF} \) is the squared error between rendered $\hat{C}(\boldsymbol{r})$ and ground truth (GT) $C(\boldsymbol{r})$ colors:

\begin{equation}
\mathcal{L}_{NeRF} = \sum_{\boldsymbol{r} \in \mathcal{R}} [\lVert \hat{C}(\boldsymbol{r}) - C(\boldsymbol{r}) \rVert^2_2],
\end{equation}
\noindent where $\mathcal{R}$ is the set of rays in each batch. \par

We describe our INGP and 3DGS, which use images captured from different camera poses as a transformation matrices. This data is typically obtained through a structure-from-motion (SfM) algorithm from CAV-frames or by transforming camera poses from the CARLA simulator to the Nerfstudio library, as demonstrated in Section \ref{sec:experimental_results}, using the following formula:
\begin{equation}
M_{ns} = R_z\left(\frac{\pi}{2}\right) R_x\left(-\frac{\pi}{2}\right) R_y\left(\pi\right) T_{\text{trans}} M_{cl},
\end{equation}

\noindent where $M_{ns}$ is a transformation matrix of the Nerfstudio library, $R_z, R_x, R_y$ represent rotation matrices for each axis, $T_{\text{trans}}$ is a transformation matrix substituting each axis and $M_{cl}$ is a transformation matrix of the CARLA simulator.

INGP \cite{mullerinstant2022} tackles the issue of NeRFs in excessive training and rendering times using neural graphics primitives and multiresolution hash encoding.
The model contains trainable weight parameters 
$\Theta$ and encoding parameters $\theta$ structured into $L$ levels, each holding up to $U$ feature vectors. Each level $l \in L$ operates independently, storing feature vectors at grid vertices. The grid resolution at each level $l$ follows a geometric progression from the coarsest $N_{min}$ to the finest $N_{max}$ resolution given by formulas:

\begin{equation}
    N_l := N_\mathrm{min} s^l,
\end{equation}

\noindent where \( N_l \) is the resolution at level \( l \), and \( s \) is the growth factor:
\begin{equation}
s :=\exp\left(\frac{\ln N_\mathrm{max}-\ln N_\mathrm{min}}{L-1}\right).
\end{equation}

The 3DGS \cite{kerbl20233d} uses differentiable 3D gaussians to model scenes without the use of neural components. 3DGS constructs 3D gaussians $G(x)$ in world space centered at point (mean) \( \mu \), and represented by covariance matrix $\Sigma$:

\begin{equation}
G(x) = exp(-\frac{1}{2} (x)^T\Sigma^{-1}(x)).
\end{equation}

These gaussians are projected to 2D using a transformation \( W \) and jacobian \( J \), resulting in a camera-space covariance matrix \( \Sigma' \):

\begin{equation}
\Sigma' = JW\Sigma W^TJ^T.
\end{equation}

The covariance matrix \( \Sigma \) defines scaling \( S_{\text{scl}} \) and rotation \( R_{\text{rot}} \):

\begin{equation}
\Sigma = R_{\text{rot}}S_{\text{scl}}S_{\text{scl}}^TR_{\text{rot}}^T.
\end{equation}

Stochastic gradient descent (SGD) optimizes the gaussian parameters \( p_{\text{gaus}}, \alpha, \Sigma \), and spherical harmonics (SH) representing color $c$ of each gaussian. The loss function \( \mathcal{L}_{3DGS} \) combines mean absolute error \( \mathcal{L}_{1} \) and a differentiable structural similarity index measure (D-SSIM) \( \mathcal{L}_{D-SSIM} \) with a balance hyperparameter \( \lambda \):

\begin{equation}
\mathcal{L}_{3DGS} = (1-\lambda)\mathcal{L}_{1} + \lambda\mathcal{L}_{D-SSIM}.
\end{equation}

\section{Formulation of the Problem}
\label{sec:problem_formulation}
In this section, we present the V2I communication problem, introducing the mathematical notation used for both the problem formulation and the proposed approach, as summarized in Table~\ref{tab:notation}. \par

\begin{figure}[!ht]
  \centering
  \includegraphics[width=1\linewidth]{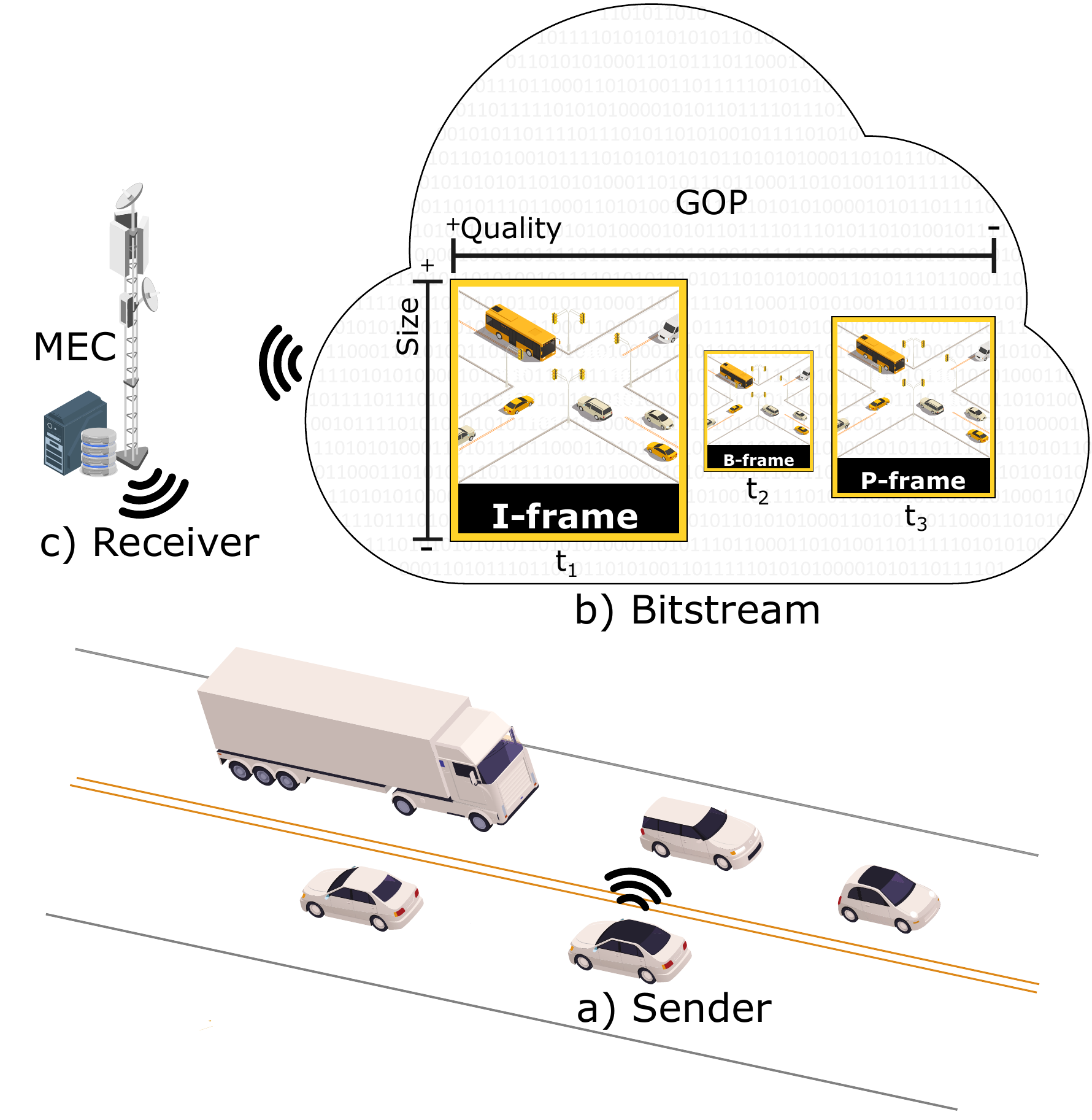}
  \caption{(a) Sender $v$ captures CAV-frames using onboard cameras. (b) A batch of CAV-frames $\mathbf{F}_{\text{cav}}$ is transmitted over the network as a bitstream $\mathbf{b}_{\text{stream}}$. (c) Receiver $e$ processes the data and returns the output to the sender.}
  \label{fig:cav_mec}
\end{figure}

\begin{table}[ht]
    \centering
    \caption{Main Notations.}
    \label{tab:notation}
    \begin{tabular}{p{3cm}|p{5cm}}
        \hline
        \textbf{Notation} & \textbf{Description} \\
        \hline
        \( \mathcal{A} \) & The set of areas \\
        \( a \) & Area \\
        \( \mathcal{V} \) & The set of CAVs \\
        \( v \) & CAV  \\
        \( \mathcal{E} \) & The set of MEC servers \\
        \( e \) & MEC server \\
        \( \mathbf{b}_{\text{stream}} \) & Bitstream \\
         \( \mathbf{b}_{\text{stream}}^{\text{ideal}} \) & Ideal bitstream \\
        \( \mathbf{F}_{\text{cav}} \) & The batch of CAV-frames \\
        \( \mathbf{F}_{\Delta} \) & The batch of Delta-frames \\
        \( \mathbf{F}_{\Delta}^{\text{ideal}} \) & The batch of ideal Delta-frames \\
        \( \mathbf{F}_{\text{rec}} \) & The batch of Rec-frames \\
        \( \Psi_{\text{rf}}\) & RF model \\
        \( \text{cam}_{\text{pose}} \) & Camera pose \\
        \( F_{\text{rf}} \) & RF-frame \\
        \( F_{\text{cav}} \) & CAV-frame \\
        \( F_{\Delta} \) & Delta-frame \\
        \( F_{\Delta}^{\text{ideal}} \) & Ideal Delta-frames \\
        \( F_{\text{rec}} \) & Rec-frame \\
        \( \mathcal{M}_{\text{rf}} \) & The set of Seg-masks from RF-frame using FastSAM\\
        \( \mathcal{M}_{\text{cav}} \) & The set of Seg-masks from CAV-frame using FastSAM \\
        \( \mathcal{M}_{\text{class}} \) & The set of Seg-masks from CAV-frame using YOLOv11 \\
        \( \mathcal{M}_{\Delta} \) & The set of Seg-masks as differences between $\mathcal{M}_{\text{rf}}$ and $\mathcal{M}_{\text{cav}}$ \\
        \( \mathcal{M}_{\Delta}^{\text{ideal}}{} \) & The ideal set of Seg-masks as differences between $\mathcal{M}_{\text{rf}}$ and $\mathcal{M}_{\text{cav}}$ \\
        \( \mathcal{S}_{\text{rf}} \) & Seg-frame from RF-frame \\
        \( \mathcal{S}_{\text{cav}} \) & Seg-frame from CAV-frame \\
        \( \mathcal{S}_{\text{class}} \) & Seg-frame from a YOLOv11 model \\
        \( \mathcal{S}_{\Delta} \) & Seg-frame with differences \\
        \( \Delta \) & Encoded delta \\
        \( \tau \) & Latency \\
        \( T_{\text{net}} \) & Available network throughput \\
        \( Q_{\text{dec}}(\mathbf{F}_{\Delta}) \) & Quality metric for decoded Delta-frames \\
        \( Q_{\text{rec}}(\mathbf{F}_{\text{rec}}) \) & Quality metric for Rec-frames \\
        \( q_{\text{min}\_\Delta} \) & Minimum required quality for Delta-frames \\
        \( q_{\text{min\_rec}} \) & Minimum required quality for Rec-frames \\
        \( \mathcal{O} \) & Set of novel objects absent in RF \\
        \( \mathcal{C}_{\text{cond}} \) & Set of environmental conditions (e.g., rain or lighting) \\
        $H$ & Codec compression efficiency \\
        \( P_{\text{loss}} \) & Packet loss probability \\
        \( r_{\text{max}} \) & Maximum compression ratio \\
        \( d_{\text{max}} \) & Theoretical maximum data savings \\
        \hline
    \end{tabular}
\end{table}

The V2I communication employs $\mathcal{V}$ as the set of CAVs and $\mathcal{E}$  as the set of MEC servers. Each CAV $v \in \mathcal{V}$ transmits batch of video frames (i.e., CAV-frames $\mathbf{F}_{\text{cav}}$) to its assigned MEC server $e \in \mathcal{E}$. Frames are grouped into batches of a specified size, denoted by $\mathbf{F}_{\text{cav}}$, and encoded before transmission. Transmitted bitstream $\mathbf{b}_{\text{stream}}$ is decoded on the receiver side for downstream tasks.

To systematically address the challenges of efficient data compression and transmission in V2I communication, we formalize the problem as an optimization task. The objective is to minimize the total amount of data transmitted while maintaining high-quality reconstruction of video frames and adhering to network constraints. This theoretical framework also helps establish an upper bound for achievable compression efficiency under ideal conditions, serving as a benchmark for evaluating the proposed RFDVC framework.

\subsection{Optimization Problem for RFDVC}

In the V2I communication scenario, a CAV captures a batch of video frames \( \mathbf{F}_{\text{cav}} \), representing its current environment. These frames are transmitted to a MEC server, where they are processed and used to synchronize P2V environments in the form of DTs. RF models \( \Psi_{\text{rf}} \) are used to model the static parts of the environment, and differences between the CAV-frames \( \mathbf{F}_{\text{cav}} \) and the RF-frames \( \mathbf{F}_{\text{rf}} \) are captured as Delta-frames \( \mathbf{F}_\Delta \).

Fig.~\ref{fig:cav_mec} illustrates a CAV $v$ (sender) offloading batch of CAV-frames $\mathbf{F}_{\text{cav}}$ over the MEC network to a nearby MEC server $e$ (receiver). A primary challenge arises from the typical use of codecs like H.264, which encode frame batches with reference and complete frames, resulting in larger data sizes and potential network congestion. Our objective is to reduce network communication overhead while ensuring that the transmitted frame batches preserve the scene semantics and information required for subsequent downstream tasks on MEC servers.

\subsubsection{Formal definitions}

   The data transmitted \( \mathbf{b}_{\text{stream}} \), is the encoded version of the Delta-frames $\mathbf{F}_\Delta$, where \( H(\cdot) \) represents the compression codec efficiency of used codec (e.g., H.264 or H.265):
   \begin{equation}
\| \mathbf{b}_{\text{stream}} \| = H (\mathbf{F}_\Delta).
   \end{equation}

\subsubsection{Objective function}   

The main goal is to minimize the total data transmitted over the network \( \mathbf{b}_{\text{stream}} \), while ensuring high-quality reconstruction \( Q_{\text{rec}}(\mathbf{F}_{\text{rec}}) \) and complying with network constraints.
The optimization problem is thus formulated as follows:
\begin{equation}
\min_{F_\Delta, \mathbf{b}_{\text{stream}}} \|\mathbf{b}_{\text{stream}}\|,
\label{eq:objective_function}
\end{equation}

subject to:
\begin{itemize}
   
    \item \ensuremath{Q_{\text{dec}}(\mathbf{F}_\Delta) \geq q_{\text{min\_$\Delta$}}, \quad \text{(see Eq.~\ref{eq:quality})}}
    \item \ensuremath{\|\mathbf{b}_{\text{stream}}\| \leq T_{\text{net}} \tau, \quad \text{(see Eq.~\ref{eq:throughput})}}
    \item \ensuremath{P_{\text{loss}}(\mathbf{b}_{\text{stream}}) \leq \epsilon, \quad \text{(see Eq.~\ref{eq:packet_loss})}}
    \item
     \ensuremath{Q_{\text{dec}}(\mathbf{F}_{\text{$\Delta$}}, \mathcal{C}_{\text{cond}}) \geq q_{\text{min\_$\Delta$}}, \quad \forall \mathcal{C}_{\text{cond}}. \quad \text{(see Eq.~\ref{eq:robustness})}}

\end{itemize}   

\subsubsection{Constraints}

The objective function is subject to the following constraints expessed by equations.

The first pair of limiting constraints in practical scenarios are the throughput and latency. Specifically, the total transmitted data \( \|\mathbf{b}_{\text{stream}}\| \) must not exceed the available network throughput \( T_{\text{net}} \) multiplied by the allowable latency \( \tau \):
   \begin{equation}
   \|\mathbf{b}_{\text{stream}}\| \leq T_{\text{net}} \tau.
   \label{eq:throughput}
   \end{equation}

To ensure that the Rec-frames \( \mathbf{F}_{\text{rec}} \) maintain fidelity, the structural similarity index measure (SSIM) or peak signal-to-noise ratio (PSNR) must meet a minimum threshold:
   \begin{equation}
   Q_{\text{rec}}(\mathbf{F}_{\text{rec}}) \geq q_{\text{min\_rec}}.
   \end{equation}

   Since the static RF component is identical for both sender and receiver and only dynamic part is transmitted through the network,  it is enough to ensure optimal reconstruction quality for Delta-frames $\mathbf{F}_\Delta$. The quality of the decoded Delta-frames \( Q_{\text{dec}}(\mathbf{F}_\Delta) \) is evaluated using SSIM. It must meet a minimum quality threshold:
   \begin{equation}
   Q_{\text{dec}}(\mathbf{F}_\Delta) \geq q_{\text{min\_$\Delta$}}.
   \label{eq:quality}
   \end{equation}

Another constraint is determined by robustness to packet loss, up to a certain limit beyond which almost no useful information can be recovered from the datastream. The impact of packet loss \( P_{\text{loss}}(\mathbf{b}_{\text{stream}}) \) must be within tolerable limits:
   \begin{equation}
   P_{\text{loss}}(\mathbf{b}_{\text{stream}}) \leq \epsilon,
   \label{eq:packet_loss}
   \end{equation}
   where \( \epsilon \) is the maximum allowable loss.

Due to existence of a wide spectrum of environments that may pose a challenge to segmentation of pixels representing dynamic  objects, the system should be robust to environmental conditions. The Rec-frames $\mathbf{F}_{\text{rec}}$ must maintain quality under these conditions \( \mathcal{C}_{\text{cond}} \), such as lighting changes, rain, or wet surfaces:
   \begin{equation}
   Q_{\text{rec}}(\mathbf{F}_{\text{rec}}, \mathcal{C}_{\text{cond}}) \geq q_{\text{min}}, \quad \forall \mathcal{C}_{\text{cond}}.
   \end{equation}

   Due to identical RF at sender and receiver we can reduce this requirement to : 

   \begin{equation}
   Q_{\text{dec}}(\mathbf{F}_{\text{$\Delta$}}, \mathcal{C}_{\text{cond}}) \geq q_{\text{min}\_\Delta}, \quad \forall \mathcal{C}_{\text{cond}}.
   \label{eq:robustness}
   \end{equation}

\subsection{Theoretical Optimum under Ideal Conditions}
Under ideal conditions, the optimization problem reaches its theoretical optimum, where:

   The ideal Delta-frame \( F_\Delta^{\text{ideal}} \) accurately captures all dynamic objects within the scene, ensuring the absence of both false positives and false negatives. It is formally defined as follows:
   \begin{equation}
   F_\Delta^{\text{ideal}}(x, y) =
\begin{cases} 
\text{Label}(M_{\Delta}^{\text{ideal}}) & \text{if pixel } (x, y) M_{\Delta}^{\text{ideal}} \in \mathcal{M}_{\Delta}^{\text{ideal}}, \\
0 & \text{otherwise (background)},
\end{cases}
\label{eq:f_ideal}
   \end{equation}
   where \( \mathcal{M}_{\Delta}^{\text{ideal}} \) is the set of ideal dynamic objects (e.g., vehicles and pedestrians) as and \( \text{Label}(M_{\Delta}^{\text{ideal}}) \) is a unique identifier assigned to each mask \( M_{\Delta}^{\text{ideal}} \) in the scene. Later in this work, in Section \ref{sec:experimental_results}, we use GT segmentation masks. This allows us to explore the gap between such ideal segmentation and non-ideal real-world segmentation approach, in terms of segmentation quality effects on compression efficiency.

   The theoretical minimum size of the transmitted bitstream \( \mathbf{b}_{\text{stream}}^{\text{ideal}} \) is:
   \begin{equation}
\| \mathbf{b}_{\text{stream}}^{\text{ideal}}\| = H(\mathbf{F}_\Delta^{\text{ideal}}),
   \end{equation}
   where \( H \) represents the compression efficiency of the codec.
   The encoded bitstream achieves the ideal maximum compression ratio \( r_{\text{max}} \):
   \[
   r_{\text{max}} = \frac{\text{RawSize}(\mathbf{F}_\Delta^{\text{ideal}})}{\text{CompSize}(\mathbf{F}_\Delta^{\text{ideal}})},
   \]
   where $\text{RawSize}(\mathbf{F}_\Delta^{\text{ideal}})$ is the size of uncompressed data and $\text{CompSize}(\mathbf{F}_\Delta^{\text{ideal}})$ is the size compressed data.\par
   Packet loss is absent in ideal conditions:
   \begin{equation}
   P_{\text{loss}} = 0.
   \end{equation}
   Thus, the decoded Delta-frames are unaffected by transmission errors. If compression is lossless as another prerequisite, we can achieve the Rec-frames $\mathbf{F}_{\text{rec}}$ achieve perfect quality:
   \begin{equation}
   Q_{\text{rec}}^{\text{ideal}}(\mathbf{F}_{\text{rec}}) = 1.
   \end{equation}

   Limited practicality of lossless compression and difficulty of driving packet loss close to zero are the main obstacles to perfect reconstruction in real world.
   The theoretical maximum data savings percentage \( d_{\text{max}} \) is given by:
   \begin{equation}
   d_{\text{max}} = 100 \left(1 - \frac{\text{CompSize}(\mathbf{F}_\Delta^{\text{ideal}})}{\text{RawSize}(\mathbf{F}_\Delta^{\text{ideal}})}\right).
   \end{equation}
Under ideal conditions, the transmitted Delta-frames \( \mathbf{F}_\Delta^{\text{ideal}} \) perfectly capture all dynamic objects, ensuring no redundancy or errors. The bitstream size is minimized to \( \mathbf{b}_{\text{stream}}^{\text{ideal}} \), determined by dynamic objects and codec efficiency. The decoded Delta-frames achieve maximum quality (\( Q_{\text{rec}}^{\text{ideal}}(\mathbf{F}_\Delta) = 1 \)), unaffected by packet loss or compression artifacts. This represents the theoretical upper bound of system performance, enabling maximum data savings and compression efficiency.

Since bitstream size is directly dependent on $\mathbf{F}_{\Delta}$ and compression efficiency of the codec used to compress $\mathbf{F}_{\Delta}$, the lower bound of bitstream size is close to 0, in case of total absence of any novel objects absent in RF, as it contains only otherwise negligible camera pose data. The upper bound is the size of $\mathbf{F}_{\text{cav}}$  compressed by such codec, in case novel objects occupy the whole field of view (FOV) of CAV camera, without any object from the static RF visible. One example of such situation in which RFDVC performance degrades that is equal to utilized base codec (H.264 or H.265) would be a vehicle in a traffic jam in front of the CAV camera covering its whole FOV.  

This theoretical framework establishes a benchmark for evaluating the practical effectiveness of the RFDVC framework and highlights the potential of leveraging RFs and Delta-frame compression under ideal conditions.

\section{Proposed Radiance Field Delta Video Compression}
Reliable communication and URLLC are critical in CAVs for transmitting high-resolution frames from onboard cameras to MEC servers for subsequent processing. In this section, we propose RFDVC, a method aimed at minimizing latency and optimizing network performance by transmitting only novel or critical data, thereby excluding redundant and repetitive information stored in distributed RFs as DTs at both ends.

\subsection{Formulation of RFDVC Problem}
\label{sub:rfdvc}

\begin{figure*}[!ht]
  \centering
  \includegraphics[width=1\linewidth]{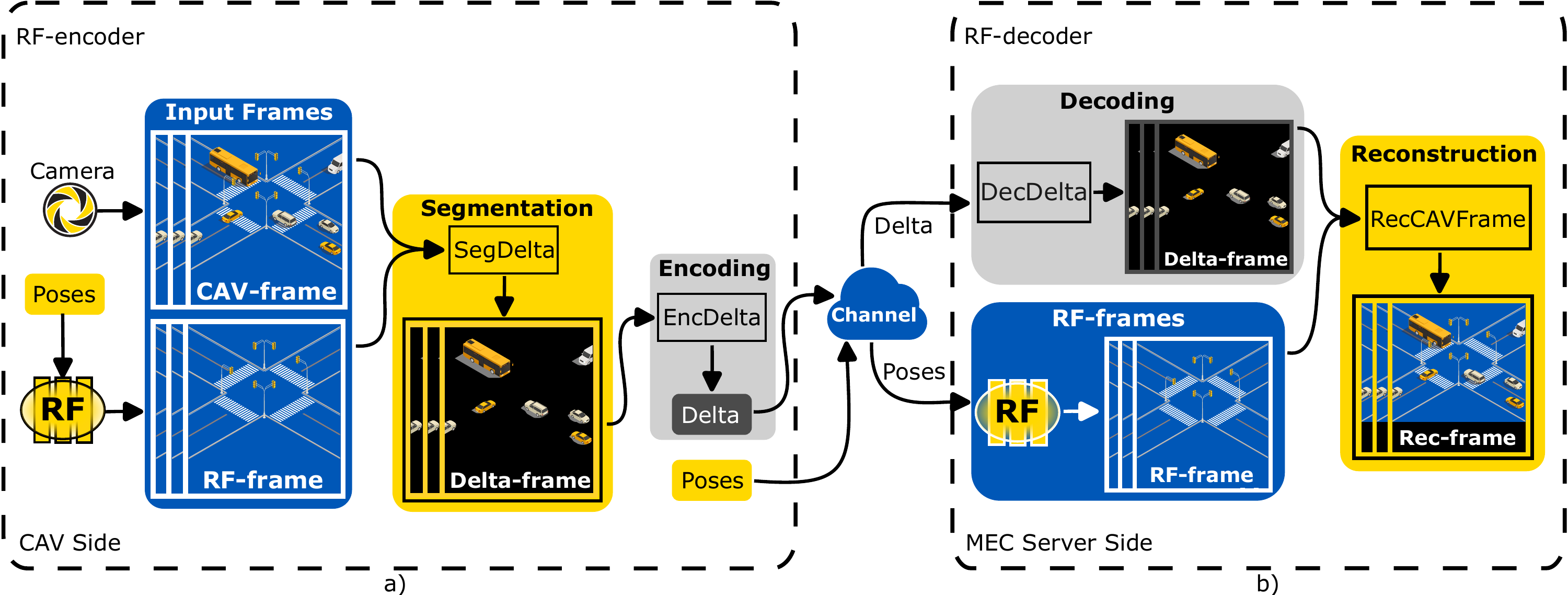}
  \caption{(a) The RF-encoder extracts pairs of input CAV-frames and RF-frames from the same camera poses to segment Delta-frames using the DS algorithm. Delta-frames are then encoded via the codec (H.264 or H.265) and transmitted over the channel. (b) The RF-decoder subsequently decodes the delta and combines Delta-frames with RF-frames to reconstruct Rec-frames.} 
  \label{fig:video_compression_rf_av}
 \end{figure*}

The proposed method employs an encoder-decoder architecture optimized for V2I communication between a CAV and an MEC server, leveraging RFs distributed at both ends. On the CAV side (sender), we use an RF-encoder as depicted in Fig.~\ref{fig:video_compression_rf_av}~a). The CAV is equipped with multiple onboard cameras that continuously capture actual traffic as CAV-frames. Concurrently, a pretrained RF model generates RF-frames capturing empty urban scene from the same camera poses without dynamic objects (e.g., vehicles or pedestrians). For each camera pose, we obtain a pair of frames: the CAV-frame and the RF-frame. \par

Our objective is to minimize the amount of data transmitted over the network. To achieve this, our RF-encoder first extract differences between image pairs, called Delta-frames. This is done using a DS algorithm to process the CAV-frame and RF-frame pairs, producing Delta-frames that are encoded using a standard video codecs. By trying to capture and separate all dynamic objects as precisely as possible, our approach strives to approach the \( \mathbf{F}_\Delta^{\text{ideal}} \) from \ref{eq:f_ideal}. This effectively allows us to get close to the global optimum of the objective function in \ref{eq:objective_function}. Details of the DS method are provided in the following Section \ref{subsec:delta_segmentation}. 
The distributed RF model represents an empty urban scene in a compressed format. In contrast to standard video codecs, which utilize I-frames (complete frames) as a primary contributor to the overall bitrate, the RF model retains this data locally, eliminating the need for its transmission over the network. This strategy substantially reduces the overall data load. \par

Fig.~\ref{fig:video_compression_rf_av}~b) illustrates the RF-decoder on the MEC server side (receiver). Using the received camera poses, RF-frames are generated with the same RF model employed at the server side. Upon receiving the encoded Delta-frames, the server decodes them and overlays these Delta-frames with the corresponding RF-frames to reconstruct the original CAV-frames, called Rec-frames. Advanced DL techniques are then applied to the Rec-frames to perform downstream tasks (e.g., object detection, semantic segmentation or advanced scene understanding). \par

\begin{algorithm}
\caption{RF-Encoder, Network Transmission,\\and RF-Decoder}
\begin{algorithmic}[1]

\Procedure{RFEncoder}{$\Psi_{\text{rf}}, \mathbf{F}_{\text{cav}}$}
    \State $\mathbf{F}_{\Delta} \gets []$
    \For{$F_{\text{cav}} \in \mathbf{F}_{\text{cav}}$}
        \State $\text{cam}_{\text{pose}} \gets \text{ExtractPose} (F_{\text{cav}})$ \Comment{Get camera pose}
        \State $F_{\text{rf}} \gets \text{RenderViewFromRF}(\Psi_{\text{rf}}, \text{cam}_{\text{pose}})$
        \State $F_{\Delta} \gets \text{SegDelta}(F_{\text{rf}}, F_{\text{cav}})$ \Comment{Extract frame differences}
        \State $\text{Append}(\mathbf{F}_{\Delta}, (\text{cam}_{\text{pose}}, F_{\Delta}))$
    \EndFor
    \State $\Delta \gets \text{EncDelta}(\mathbf{F}_{\Delta})$ \Comment{Encode Delta-frames}
    \State \textbf{return} $\Delta$
\EndProcedure
\label{alg:rf_encoder}

\vspace{1em} 

\Procedure{ChannelTX}{$\Delta, T_{\text{net}}$}
    \State $\mathbf{b}_{stream} \gets []$
    \State $\text{size}_{\Delta} \gets \text{GetSize}(\Delta)$
    \State $\tau \gets \text{size}_{\Delta} / T_{\text{net}}$ \Comment{Estimate transmission time}
    \State $\text{TransmitOverChannel}(\Delta, \tau)$
    \State $\text{Append}(\mathbf{b}_{stream}, (\Delta, \tau))$
    \State \textbf{return} $\mathbf{b}_{stream}$
\EndProcedure
\label{alg:network_tx}

\vspace{1em} 

\Procedure{RFDecoder}{$\mathbf{b}_{stream}, \Psi_{\text{rf}}$}
    \State $\mathbf{F}_{rec} \gets []$
    \State ($\Delta, \tau) \gets \text{GetData}(\mathbf{b}_{stream})$
    \State $\text{WaitForTransmission}(\tau)$
    \State $\mathbf{F}_{\Delta} \gets \text{DecDelta}(\Delta)$ \Comment{Decode Delta-frames}
    \For{each $(\text{cam}_{\text{pose}}, F_{\Delta}) \in \mathbf{F}_{\Delta}$}
        \State $F_{\text{rf}} \gets \text{RenderViewFromRF}(\Psi_{\text{rf}}, \text{cam}_{\text{pose}})$\\ \Comment{Render static view}
        \State $F_{\text{rec}} \gets \text{RecCAVFrame}(F_{\text{rf}}, F_{\Delta})$ \Comment{Reconstruct Rec-frame}
        \State $\text{Append}(\mathbf{F}_{rec}, F_{\text{rec}})$
    \EndFor
    \State \textbf{return} $\mathbf{F}_{rec}$
\EndProcedure
\label{alg:rf_decoder}

\end{algorithmic}
\label{alg:pseudocode_rfdvc}
\end{algorithm}

Alg.~\ref{alg:pseudocode_rfdvc} comprises three core procedures \texttt{RFEncoder}, \texttt{ChannelTX}, and \texttt{RFDecoder} designed to optimize frame encoding, network transmission, and decoding using RFs. In the \texttt{RFEncoder} procedure, lines $1\text{-}11$ iterate over each frame in \texttt{CAV\_Frames}, where the camera pose is extracted, RF-frame is rendered based on the pose, and the Delta-frame is computed by comparing CAV-frame and RF-frame. These computed Delta-frames are accumulated in a list and subsequently encoded. The \texttt{ChannelTX} procedure, detailed in lines $12\text{-}19$, segments the encoded deltas for transmission according to the available network throughput, with realistic packet loss. Finally, the \texttt{RFDecoder} procedure, represented by lines $20\text{-}31$, receives the transmitted deltas, decodes them, and reconstructs each frame by merging the decoded Delta-frame with a freshly rendered RF-frame aligned to the transmitted pose, preserving frame integrity upon reception. This three stage process reduces bandwidth by transmitting only delta information, thereby enabling efficient, high-quality frame reconstruction on the receiver side.

\subsection{Delta Segmentation}
\label{subsec:delta_segmentation}
The DS algorithm extracts differences between two images: the CAV-frame \(F_{\text{cav}}\) captured by the vehicle’s onboard camera and the RF-frame \(F_{\text{rf}}\) generated by the RF model. This process focuses on isolating differences, such as vehicles and pedestrians, omitting static components (e.g., buildings, roads or traffic lights), which are already modeled by the RF. By substituting irrelevant regions with black pixels, DS improves the compression of data transmitted to the MEC server. \par

We use the segment anything model (SAM) \cite{kirillow2023sam} for its zero-shot generalization, enabling robust segmentation without being limited by predefined classes. However, due to its high computational requirements, we employ FastSAM \cite{zhao2023fastsam}, a faster convolutional neural network (CNN)-based variant with comparable accuracy, for real-time applications.

\begin{algorithm}
\caption{Delta Segmentation}
\begin{algorithmic}[1]
\Procedure{SegDelta}{$F_{\text{rf}}, F_{\text{cav}}$}

\State $\mathcal{M}_{\text{rf}} \gets \text{FastSAM} (F_{\text{rf}})$
\State $\mathcal{M}_{\text{cav}} \gets \text{FastSAM} (F_{\text{cav}})$
\State $\mathcal{M}_{\Delta} \gets \emptyset$
\For{$M_{\text{rf}} \in \mathcal{M}_{\text{rf}}$}
    \For{$M_{\text{cav}} \in \mathcal{M}_{\text{cav}}$}
        \State $M_\Delta \gets \text{ComputeIoU}(M_{\text{rf}}, M_{\text{cav}})$
            \State $\mathcal{M}_\Delta \gets \mathcal{M}_\Delta \cup \{M_\Delta\}$
    \EndFor
\EndFor
\State $S_{\Delta} \gets \text{SegFrameFromMasks} 
(\mathcal{M}_{\Delta})$
\State $\mathcal{M}_{\text{class}} \gets \text{YOLOv11} 
(F_{\text{cav}})$
\State $S_{\text{class}} \gets \text{SegFrameFromMasks} 
(\mathcal{M}_{\text{class}})$
\State $F_{\Delta} \gets \text{Overlay} 
(S_\Delta, S_{\text{class}})$
\State \textbf{return} $F_{\Delta}$
\EndProcedure

\end{algorithmic}
\label{alg:delta_segmentation}
\end{algorithm}

In this part, we provide a thorough  description of the DS algorithm, which is presented in Alg. \ref{alg:delta_segmentation}. The algorithm is encapsulated within the single function \texttt{SegDelta}, which serves as the core procedure for implementing the DS algorithm.  Lines $2\text{-}3$ utilize FastSAM to segment both the RF-frame \(F_{\text{rf}}\) and the CAV-frame \(F_{\text{cav}}\) resulting in the corresponding segmentaiton masks (Seg-masks) $\mathcal{M}_{\text{rf}}$ and $\mathcal{M}_{\text{cav}}$, respectively. In line $4$, an empty set of masks $\mathcal{M}_\Delta$ is initialized to store the masks considered as differences. Lines $5\text{-}10$ compare the masks using the function \texttt{ComputeIoU}, which employs the intersection-over-union (IoU) metric, defined as follows:
\begin{equation}
\text{IoU}(M_{\text{rf}}, M_{\text{cav}}) = \frac{|M_{\text{rf}} \cap M_{\text{cav}}|}{|M_{\text{rf}} \cup M_{\text{cav}}|},
\end{equation}
\noindent where \(| \cdot |\) represents the area of a given mask.

Masks are considered matching if:
\begin{equation}
\text{IoU}(M_{\text{rf}}, M_{\text{cav}}) > \delta_{\text{thr}},
\end{equation}
\noindent where \(\delta_{\text{thr}}\) is a predefined threshold. \par

Masks that do not satisfy the criterion \(\text{IoU}(M_{\text{rf}}, M_{\text{cav}}) \leq \delta_{\text{thr}}\) are identified as differences $M_\Delta$ and are subsequently added to the set \(\mathcal{M}_{\Delta}\). In line $11$, the function \texttt{SegFrameFromMasks} is employed to construct segmentation frame (Seg-Frame)  $S_\Delta$, which is composed of the masks contained in $\mathcal{M}_\Delta$.

To ensure reliable segmentation of critical object classes (e.g., pedestrians and vehicles) for transmission and analysis on the MEC server, we employ the YOLOv11 model \cite{khanam2024yolov11}. The YOLOv11x variant delivers high segmentation accuracy, achieving approximately$54.5\%$ $\text{mAP}^{\text{50–95}}$ with a latency of 13 ms. In line $12$, the YOLOv11 model \cite{khanam2024yolov11}   is applied to the CAV-frame, resulting in the generation of a set of masks $\mathcal{M}_{\text{class}}$ corresponding to the critical object classes. Subsequently, in line $13$, the set of masks $\mathcal{M}_{\text{class}}$ is utilized to construct the Seg-frame \(S_{\text{class}}\) using the \texttt{SegFrameFromMask} function.\par

\begin{figure}[!ht]
  \centering
  \includegraphics[width=0.6\linewidth]{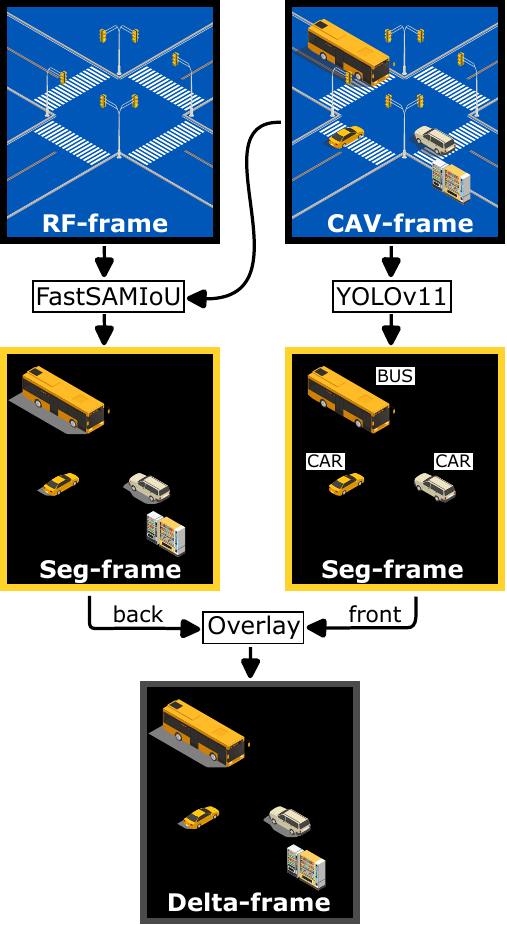}
  \caption{Differences between the CAV-frame and RF-frame are segmented using FastSAM for its zero-shot generalization and IoU metrics, while YOLOv11 ensures segmentation of critical classes (e.g., pedestrians and vehicles). The Delta-frame overlays the YOLO masks onto the FastSAM masks, with irrelevant pixels represented in black.}
  \label{fig:fastsam_yolo}
\end{figure}

In line $14$, the Delta-frame \( F_{\Delta} \) is generated by compositing 
\( S_{\Delta} \) and \( S_{\text{class}} \) by  overlaying them using the \texttt{Overlay} function, while masking irrelevant regions with black pixels. The overall process is illustrated in Fig.~\ref{fig:fastsam_yolo}, where the \texttt{FastSAMIoU} function encapsulates
lines $2\text{-}11$ for improved clarity. \par

\section{Experimental Results}
\label{sec:experimental_results}
An experimental setup was established in a virtual urban environment using the CARLA simulator to validate this approach. This setup aimed to replicate real-world urban scenarios within a controlled, simulated environment, providing a robust platform for testing the efficacy of the RF-based compression system in an urban CAV context.\par
The simulation results of the RFDVC approach are obtained using simulation code that we have made available in an online GitHub repository\footnote{\url{https://github.com/Maftej/rfdvc}}.

\subsection{Simulation Parameters}
To ensure reproducibility, we provide key details on the simulation environment, dataset used, and model configurations.
Table \ref{tab:parameters_carla_sim} outlines the parameters used in the CARLA simulator. Specifically, we used CARLA version 0.9.15 with the Town10 map. Nine cameras were employed during training, while three were used during deployment, with all cameras capturing images at a resolution of 1920x1080 pixels. The motion blur intensity was set to zero, the focal distance was 1000 cm, and the FOV was 90$^{\circ}$.

\begin{table}[ht]
\centering
\caption{Parameters of CARLA simulator.}
\label{tab:parameters_carla_sim}
\begin{tabular}{ll}
\hline
\textbf{Parameter} & \textbf{Value} \\ \hline
CARLA version & 0.9.15 \\
Map & Town10 \\
Cameras during training / deployment & 9 / 3 \\
Resolution of cameras \& images & 1920x1080 \\
Motion blur intensity & 0 \\
Focal distance & 1000 cm \\
FOV & 90$^{\circ}$ \\
\hline
\end{tabular}
\end{table}

\begin{figure*}[!ht]
  \centering
  \includegraphics[width=1\linewidth]{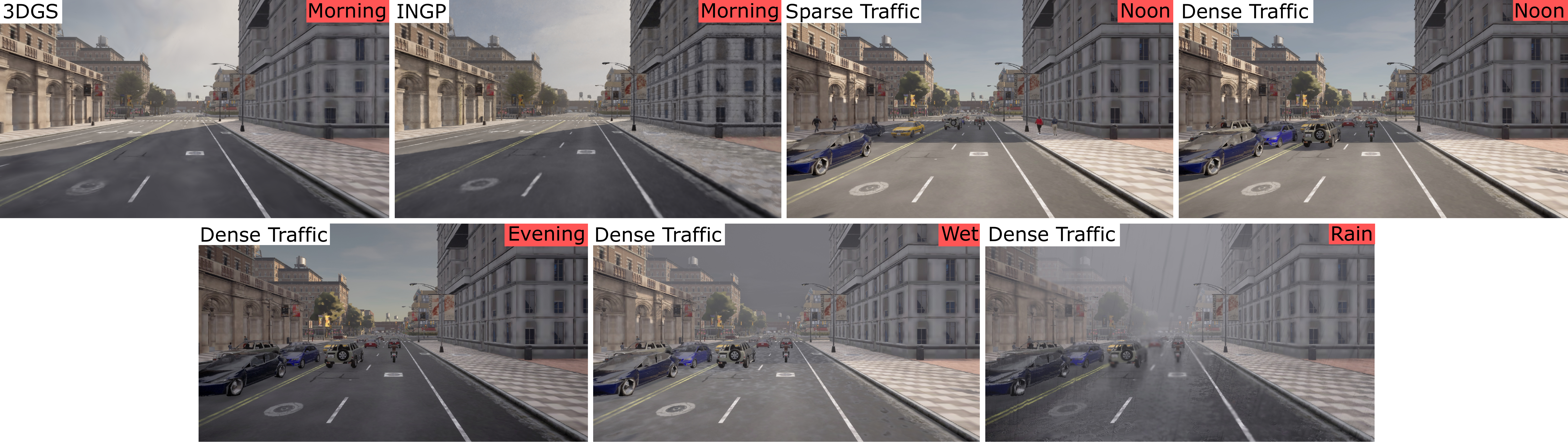}
  \caption{Examples of RF-frames extracted from 3DGS and INGP RF models trained in morning conditions with no vehicles or pedestrians present, followed by CAV-frames from sparse traffic in noon and dense traffic in noon, evening, wet conditions, and rain.}
  \label{fig:weather_conditions}
\end{figure*}

Table \ref{tab:dataset_default_weather} details the datasets utilized for training and deployment under morning conditions. The training dataset was collected in a static environment, with no dynamic or temporary phenomena like traffic, pedestrians, or vehicles. For deployment, two scenarios were considered: sparse traffic with 60 pedestrians and 100 vehicles, and dense traffic with 60 pedestrians and 200 vehicles. Fig.~\ref{fig:weather_conditions} illustrates the diverse conditions and traffic density as generated by the CARLA simulator. The morning conditions scenario is employed for training the RF models 3DGS and INGP. Conditions at noon, evening, and during wet conditions are closely aligned with the morning scenario, making the generation of Delta-frames relatively straightforward. In contrast, rain conditions differ significantly from the morning conditions, presenting greater challenges in the production of Delta-frames. 

\begin{table}[ht]
\centering
\caption{Dataset for training and deployment.}
\label{tab:dataset_default_weather}
\begin{tabular}{lllll}
\hline
\textbf{Dataset} & \textbf{Traffic} & \textbf{Pedestrians} & \textbf{Vehicles} \\ \hline
Training & Empty & 0 & 0 \\
Deployment & Sparse & 60 & 100 \\
Deployment & Dense & 60 & 200  \\ \hline
\end{tabular}
\end{table}

We employed RF models INGP and 3DGS using the nerfstudio library, configured with the FOV and image resolution specified in Table \ref{tab:parameters_carla_sim}. Both models were trained for 30 000 iterations.

\subsection{Quantitative Evaluation of RF Models}
To assess potential quality degradation introduced by RFs and the impact of environmental changes, we employed key image quality metrics: PSNR, SSIM, and learned perceptual image patch similarity (LPIPS). These metrics evaluate differences in pixel values, structural features, and perceptual quality, respectively.\par
\begin{figure}[!ht]
  \centering
  \includegraphics[width=0.8\linewidth]{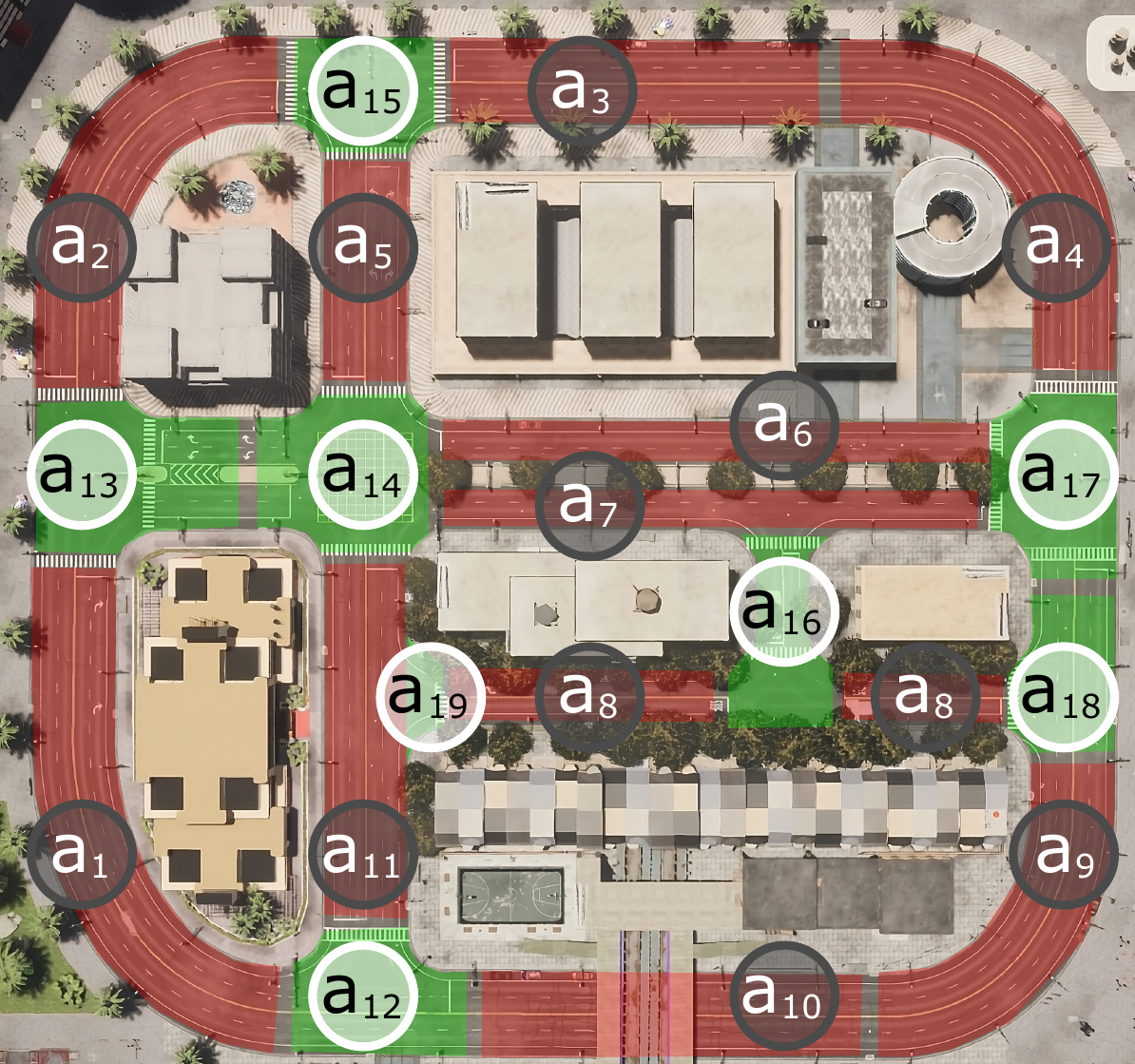}
  \caption{CARLA map divided into 19 areas, each represented as RF model. Highways are depicted in red and intersections in green.}
  \label{fig:carla_map_segments}
\end{figure}
V2I communication is implemented within the city map. Given the limitations of RFs to smaller scenes, we segment the city's road network into distinct areas $a \in \mathcal{A}$, as depicted in Fig.~\ref{fig:carla_map_segments}. The map is color-coded, with highways represented in red and intersections in green. In total, the map consists of 19 areas, each designed to correspond to specific regions of interest within the simulation environment.

\begin{table}[ht]
\centering
\caption{Quantitative evaluation of INGP and 3DGS models.}
\label{tab:evaluation_rf_models}
\resizebox{\columnwidth}{!}{%
\begin{tabular}{|l|lll|lll|lllll}
\hline
 & \multicolumn{3}{c|}{\textbf{INGP}} & \multicolumn{3}{c|}{\textbf{3DGS}} \\ \cline{2-7} 
\textbf{Area} & \textbf{PSNR} $\uparrow$ & \textbf{SSIM} $\uparrow$ & \textbf{LPIPS} $\downarrow$ & \textbf{PSNR} $\uparrow$ & \textbf{SSIM} $\uparrow$ & \textbf{LPIPS} $\downarrow$ \\ \hline
$a_1$ & 20.68 & 0.65 & 0.53 & 25.17 & 0.78 & 0.41 \\
$a_2$ & 19.99 & 0.58 & 0.52 & 25.37 & 0.78 & 0.41 \\
$a_3$ & 21.25 & 0.63 & 0.49 & 28.35 & 0.84 & 0.34 \\
$a_4$ & 19.38 & 0.56 & 0.53 & 25.20 & 0.77 & 0.41 \\
$a_5$ & 21.10 & 0.64 & 0.47 & \cellcolor{LightGreen}28.94 & \cellcolor{LightGreen}0.86 & \cellcolor{LightGreen}0.32 \\
$a_6$ & 20.81 & 0.58 & 0.53 & 27.03 & 0.81 & 0.36 \\
$a_7$ & 20.52 & 0.58 & 0.52 & 27.15 & 0.82 & 0.35 \\
$a_8$ & 20.87 & \cellcolor{LightRed}0.50 & \cellcolor{LightRed}0.61 & 26.62 & 0.72 & 0.45 \\
$a_9$ & 20.34 & 0.60 & 0.53 & 25.70 & 0.78 & 0.41 \\
$a_{10}$ & 19.85 & 0.55 & 0.55 & 25.77 & 0.81 & 0.38 \\
$a_{11}$ & 21.91 & 0.65 & 0.51 & 27.98 & 0.82 & 0.37 \\
$a_{12}$ & 21.38 & 0.67 & 0.48 & 26.14 & 0.82 & 0.38 \\
$a_{13}$ & 19.41 & 0.58 & 0.51 & 25.54 & 0.80 & 0.39 \\
$a_{14}$ & \cellcolor{LightRed}18.99 & 0.54 & 0.52 & \cellcolor{LightRed}23.58 & 0.73 & 0.44 \\
$a_{15}$ & 20.12 & 0.51 & 0.50 & 24.81 & \cellcolor{LightRed}0.71 & 0.39 \\
$a_{16}$ & 19.36 & 0.56 & 0.56 & 25.01 & 0.78 & \cellcolor{LightRed}0.46 \\
$a_{17}$ & 20.71 & 0.65 & 0.48 & 26.13 & 0.80 & 0.37 \\
$a_{18}$ & \cellcolor{LightGreen}22.44 & \cellcolor{LightGreen}0.69 & \cellcolor{LightGreen}0.47 & 27.20 & 0.82 & 0.37 \\
$a_{19}$ & 19.99 & 0.52 & 0.51 & 28.18 & 0.84 & 0.34 \\ \hline
\end{tabular}%
}
\end{table}

Table \ref{tab:evaluation_rf_models} presents the average PSNR, SSIM, and LPIPS values for both INGP and 3DGS models across various areas. In the INGP models, area $a_{18}$ achieved the highest PSNR of 22.44, indicating better image fidelity, while area $a_{14}$ recorded the lowest PSNR of 18.99, suggesting degradation. SSIM values were highest in area $a_{12}$ and lowest in area $a_{10}$. For LPIPS, area $a_5$ performed best, whereas area $a_8$ exhibited the highest perceptual dissimilarity.\par
In comparison, the 3DGS models generally showed higher PSNR and SSIM values, with area $a_5$ excelling in both PSNR 28.94 and LPIPS 0.32. However, area $a_{14}$ underperformed, demonstrating the lowest PSNR and higher LPIPS, indicating potential issues in perceptual quality.

Building on these findings, all experiments in this paper leverage the 3DGS model trained on area $a_5$, as it consistently delivered the best results across key image quality metrics. By utilizing this model, we focus on showcasing the optimal performance of the RFDVC framework under various experimental conditions. Future work could investigate the use of models trained on multiple areas to evaluate the generalizability of the approach across diverse areas.

\subsection{Data Savings under Different Conditions}
A primary feature of RFDVC is its substantial data savings compared to the commonly used and highly efficient H.264 and H.265 codecs. For URLLC, frames are transmitted in batches of up to 10 at 30 FPS to support real-time communication. Images are encoded with consistent configurations across both H.264 and RFDVC, with the latter leveraging adaptive quantization to more effectively compress black regions containing no information relevant to downstream tasks. Furthermore, we evaluate Delta-frame encoding using GT masks as an upper bound for compression efficiency against our DS algorithm, which offers a realistic encoding approach to capture most differences without restrictions to class categories, as in methods such as YOLO or other DL models used in computer vision tasks.\par
\begin{figure*}[!ht]
  \centering
\includegraphics[width=0.65\linewidth]{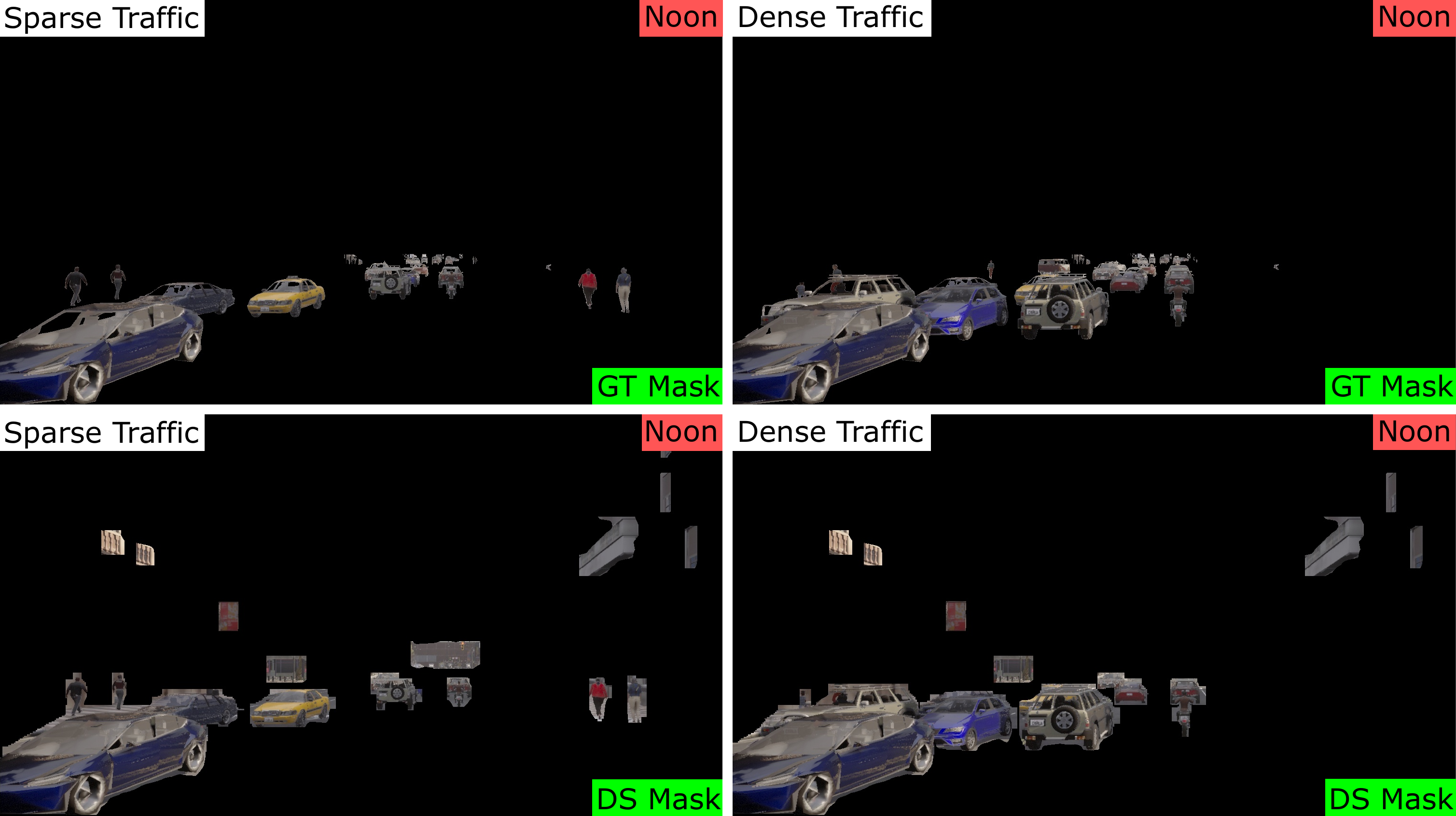}
  \caption{Delta-frames extracted using GT masks from CARLA simulator and using proposed DS algorithm in both sparse and dense traffic in noon conditions. Artifacts introduced by the DS algorithm lead to an increase in data transmission requirements.}
  \label{fig:df_gtms}
\end{figure*}
The simulation environment facilitates the extraction of both RGB images and GT segmentation masks, which are then processed to retain only vehicles and pedestrians, with the remaining masked areas of the image represented by black pixels. It is assumed that the RF models used in this process exhibit no discrepancies and that the environment contains no additional objects. Fig.~\ref{fig:df_gtms} depicts both the Delta-frames extracted from the GT masks and proposed DS algorithm in sparse and dense traffic. DS method allows for the extraction of all relevant information necessary for downstream tasks without requiring knowledge of the exact object class.\par
\begin{figure}[!ht]
  \centering
  \includegraphics[width=1\linewidth]{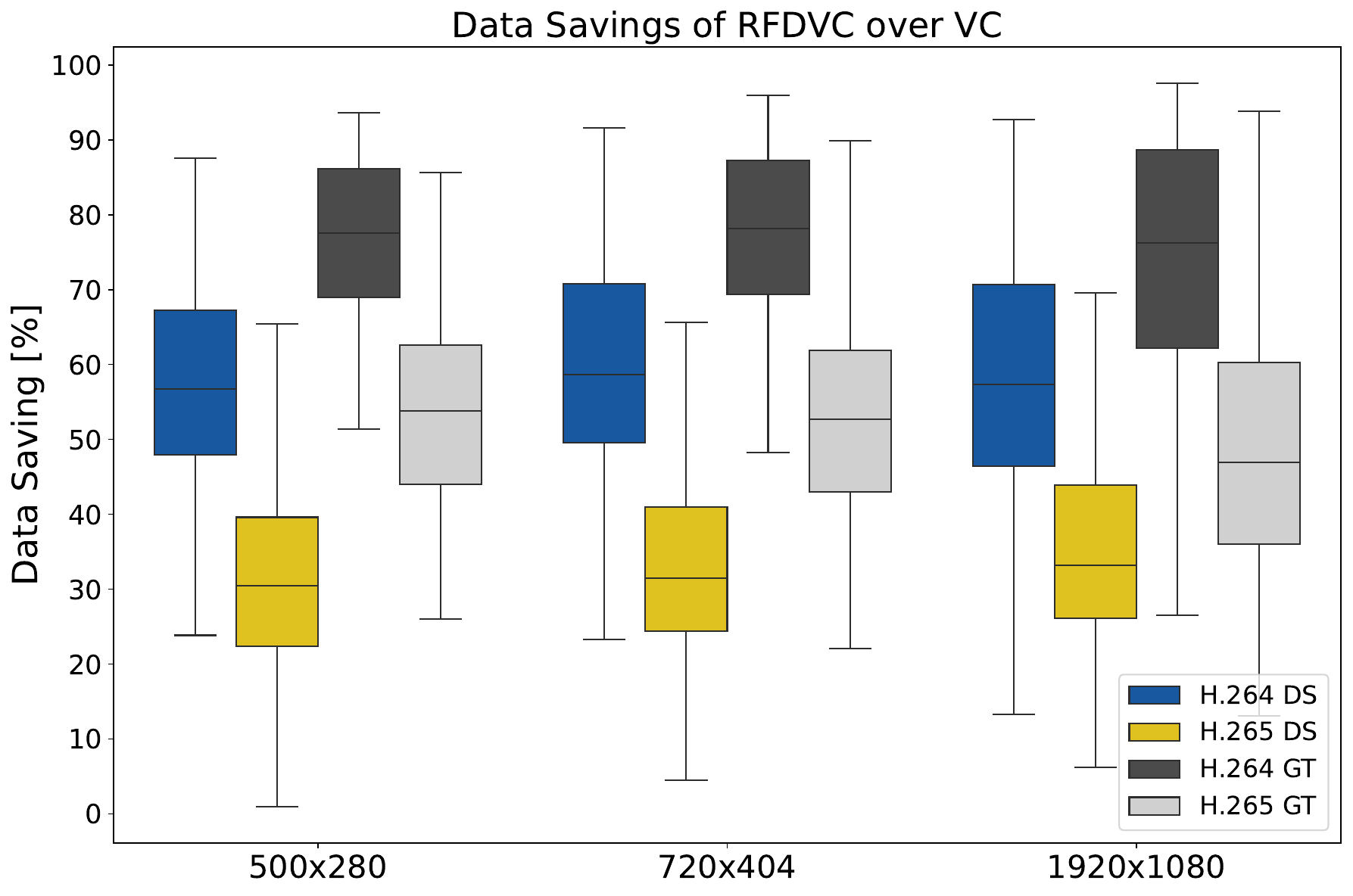}
  \caption{RFDVC data savings, for both H.264 and H.265-based RFDVC variants, utilizing masks obtained using DS method and GT masks in noon, evening and wet conditions. H.264-based RFDVC savings are measured relative to plain H.264 maskless frame compression, while H.265-based RFDVC savings are measured relative to plain H.265 codec.}
\label{fig:rfdvc_a5_3dgs_noon_evening_wet}
\end{figure}
Fig.~\ref{fig:rfdvc_a5_3dgs_noon_evening_wet} displays a box plot comparing the data savings achieved by RFDVC and VC when using either the DS algorithm or GT masks for transmitting data from camera sensors under noon, evening, and wet conditions. While static RF models are applied, the conditions within these models do not align precisely with actual conditions, resulting in lighting discrepancies, particularly under wet conditions scenarios. For RFDVC with the DS algorithm, the interquartile range of data savings spans 48\% to 71\% with the H.264 codec and 24\% to 44\% with the H.265 codec. Higher resolutions provide greater data savings as larger black regions are more effectively encoded. With GT masks, data savings vary from approximately 65\% to 90\% using the H.264 codec and from 38\% to 63\% using the H.265 codec.\par
\begin{figure}[!ht]
  \centering
  \includegraphics[width=1\linewidth]{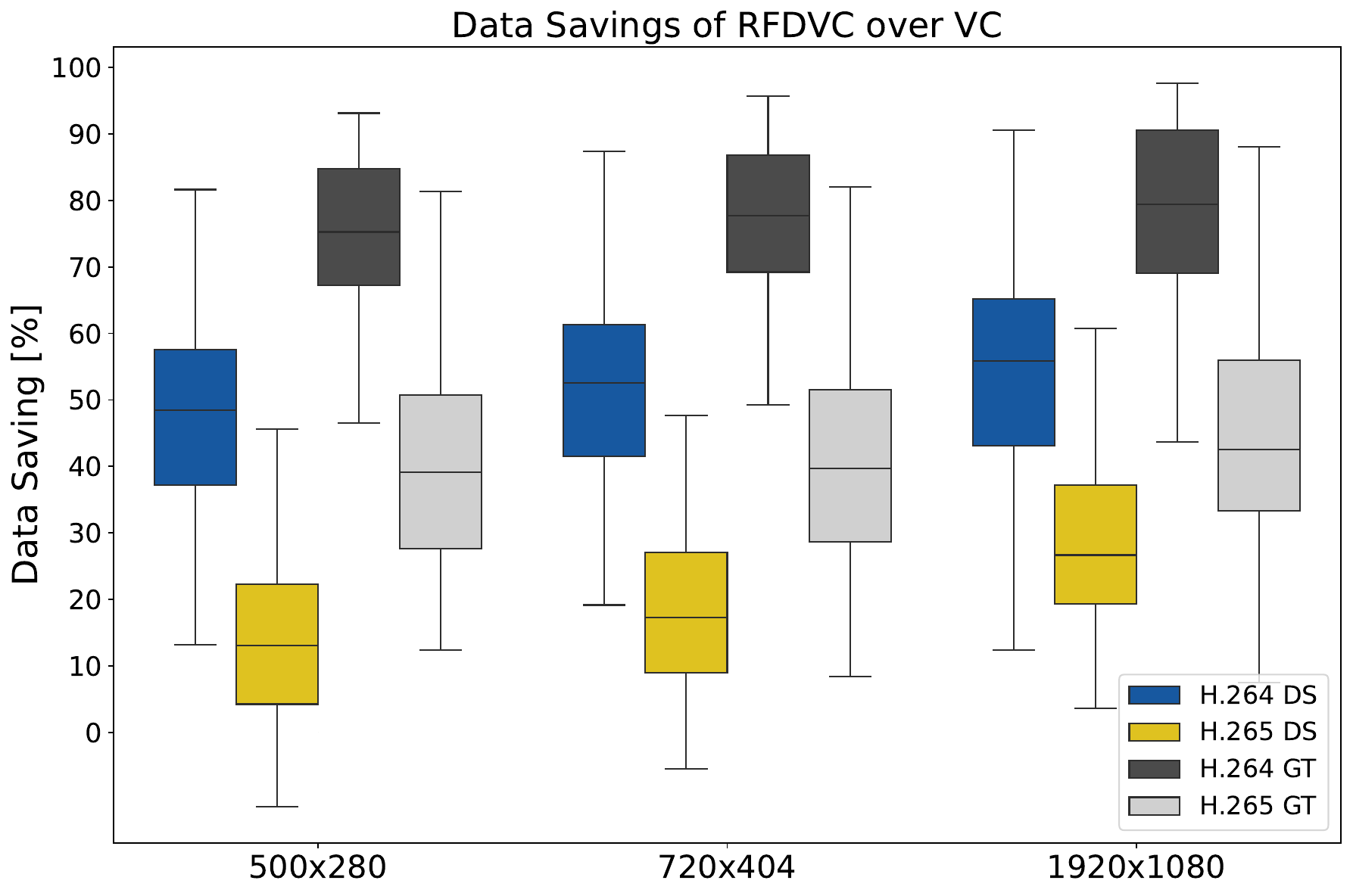}
  \caption{RFDVC data savings, for both H.264 and H.265-based RFDVC variants, utilizing masks obtained using DS method and GT masks in rainy weather conditions. H.264-based RFDVC savings are measured relative to plain H.264 maskless frame compression, while H.265-based RFDVC savings are measured relative to plain H.265 codec.}
  \label{fig:rfdvc_a5_3dgs_rain}
\end{figure}
Fig.~\ref{fig:rfdvc_a5_3dgs_rain} illustrates the data savings achieved by RFDVC compared to VC under rain conditions, which pose a more challenging environment due to substantial lighting changes and partial frame corruption from heavy rainfall. Despite these challenging conditions, data savings fall within an interquartile range of 39\% to 68\% with the H.264 codec and 7\% to 37\% with the H.265 codec. Using GT masks, data savings are approximately 68\% to 90\% with the H.264 codec and 29\% to 59\% with the H.265 codec.

\subsection{Packet Loss}
\begin{figure}[!ht]
  \centering
  \includegraphics[width=0.8\linewidth]{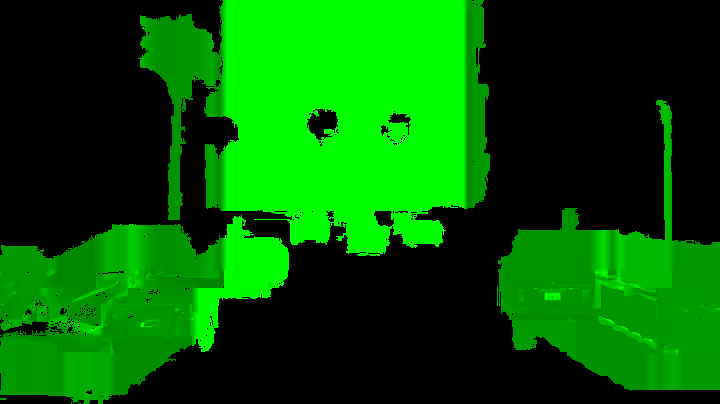}
  \caption{A decoded Delta-frame exhibiting corruption caused by simulated packet loss. The background, represented by black pixels, remains unchanged as the positions of corrupted masks are transmitted in the form of polygons.}
  \label{fig:corrupted_frame}
\end{figure}

We simulate packet loss using the Gilbert-Elliott model \cite{gilbert1960capacity}, a stochastic model widely employed to represent bursty error patterns in communication channels. The model alternates between "good" and "bad" states, capturing the temporal correlation of packet losses with distinct probabilities for each state. We demonstrate that the RFDVC approach surpasses standard H.264 and H.265 codecs, even under packet loss conditions. The RFDVC method shows greater effectiveness in ensuring that all critical information is reliably transmitted across the network, a crucial requirement for real-time and sensitive communication. This advantage is especially significant in the context of CAVs, where safety and reliability must be maintained across diverse operational scenarios. Polygon representation can be effectively employed to facilitate the efficient transmission of mask positions within the frames, enabling the exclusion of corrupted regions associated with the background from the Rec-frame, as illustrated in Fig.~\ref{fig:corrupted_frame}.\par

\begin{figure}[!ht]
  \centering
  \includegraphics[width=1\linewidth]{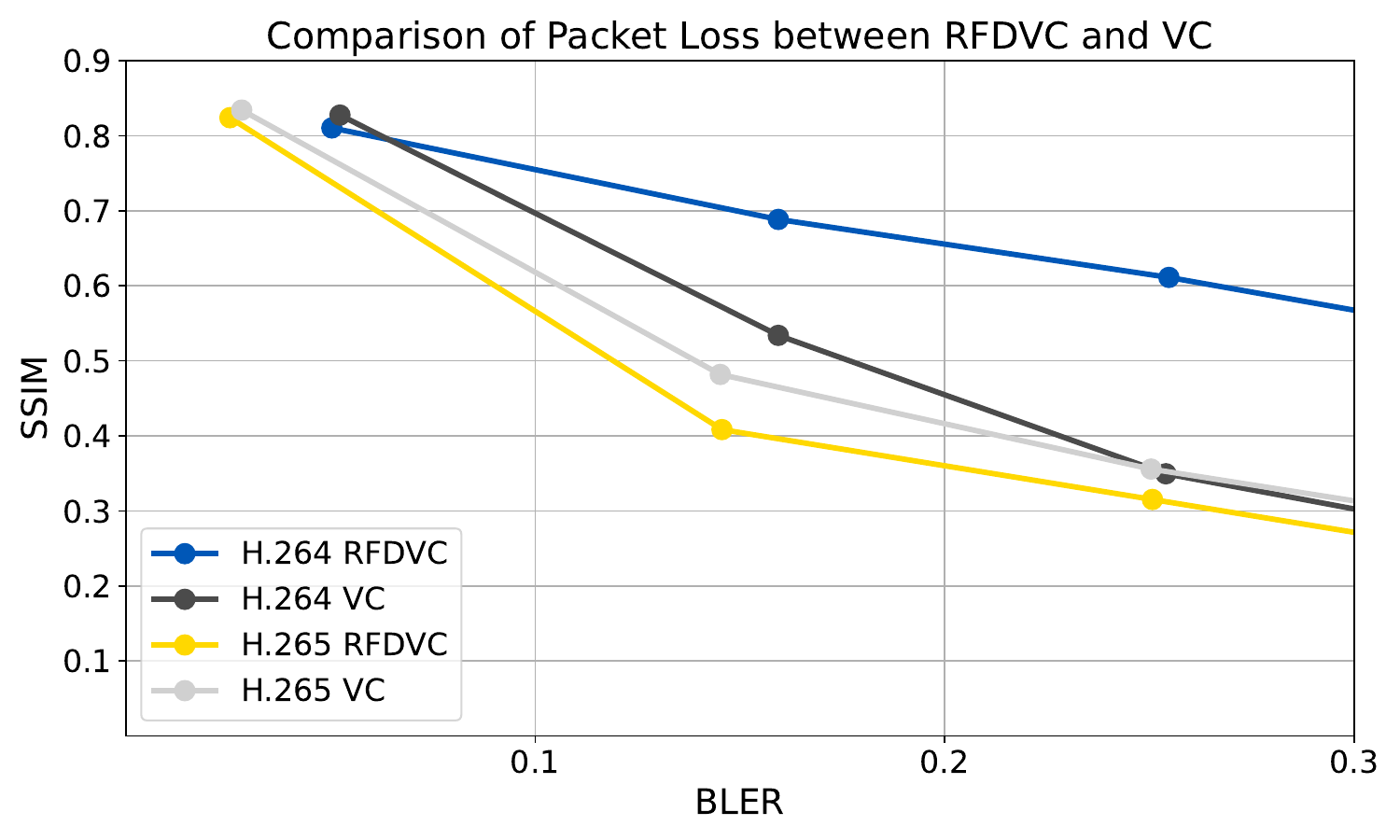}
  \caption{Influence of a simulated packet loss on frame reconstruction quality (SSIM) for RFDVC and VC approaches, excluding rainy conditions, evaluated across multiple resolutions and simulated environments. The results compare encoding using H.264 and H.265 codecs leveraging RFDVC against their standalone implementations, emphasizing static and wet conditions without rainfall.}
  \label{fig:ssim_bler_3weathers}
\end{figure}

\begin{figure}[!ht]
  \centering
  \includegraphics[width=1\linewidth]{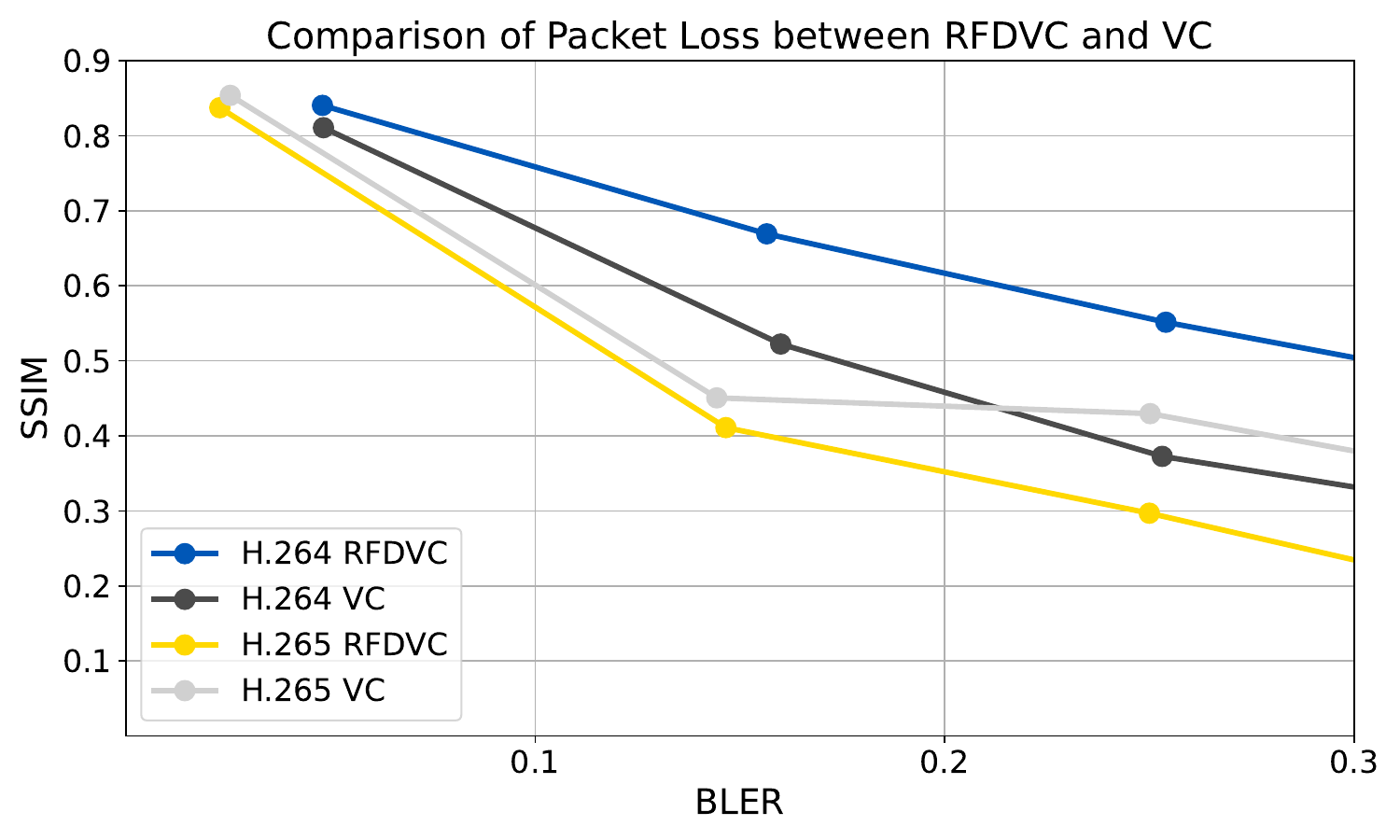}
  \caption{Impact of rainy weather on frame reconstruction quality (SSIM) under simulated packet loss for RFDVC and VC approaches. The evaluation includes encoding with H.264 and H.265 codecs and highlights the performance resilience of RFDVC under challenging conditions caused by rainfall.}
  \label{fig:ssim_bler_rain}
\end{figure}

Figs.~\ref{fig:ssim_bler_3weathers} and \ref{fig:ssim_bler_rain} depict the impact of block error rate (BLER) on the SSIM of reconstructed video frames, comparing two RFDVC variants (H.264 and H.265) against their standard VC counterparts, under different environmental conditions.

Fig.~\ref{fig:ssim_bler_3weathers} shows results averaged across all conditions and resolutions, excluding rainy weather. Here, the H.264-based RFDVC exhibits a significantly larger gap in SSIM performance compared to other approaches, consistently achieving superior reconstruction quality as packet loss increases. This advantage is particularly evident at a \textbf{BLER of approximately 0.25}, where the H.264 RFDVC achieves an SSIM of $\sim0.61$. This significantly outperforms the standard H.264 VC approach, which records an SSIM of $\sim0.35$ (\textbf{+0.26} improvement). Similarly, at this error rate, H.264 RFDVC surpasses H.265 VC ($\sim0.36$, \textbf{+0.25}) and H.265 RFDVC ($\sim0.32$, \textbf{+0.29}). These results highlight RFDVC's resilience, especially for the H.264-based implementation, in delivering higher reconstruction quality under diverse conditions.\par

Fig.~\ref{fig:ssim_bler_rain} isolates the analysis to rainy weather conditions, showing a more pronounced SSIM degradation across all methods compared to the average results in Fig.~\ref{fig:ssim_bler_3weathers}. The H.264-based RFDVC still demonstrates a clear edge in quality over its standard VC counterpart. At a \textbf{BLER of approximately 0.25}, H.264 RFDVC achieves an SSIM of $\sim0.55$, which is substantially higher than the standard H.264 VC's SSIM of $\sim0.37$ (\textbf{+0.18} improvement). However, the challenging conditions of heavy rain diminish the effectiveness of the H.265-based RFDVC, which at the same BLER ($\sim0.30$ SSIM) underperforms relative to the standard H.265 VC ($\sim0.43$ SSIM). This suggests that severe visual artifacts from rain can complicate the delta segmentation process, making the standard codec more effective in this specific scenario.

These findings affirm that the H.264-based RFDVC consistently delivers superior performance across all tested conditions. In contrast, while the more efficient H.265 RFDVC performs well with low packet loss, its visual quality degrades significantly as the error rate increases, underperforming relative to standard VC methods at BLER values above approximately 0.1 in both rainy and non-rainy conditions. This can be explained by the inherent nature of H.265, which contains less redundant information to achieve larger compression ratios, diminishing its resilience to packet damage.

\section{Conclusions}
This work proposes an RFDVC approach, based on RF-encoder and RF-decoder architecture, for V2I communication and P2V synchronization within the vehicular metaverse. Distributed RFs serve as DTs storing photorealistic 3D scene in compressed form. An experimental setup was established in a virtual urban environment using the CARLA simulator divided into 19 areas represented as RFs. The experimental results show that RFDVC achieves significant data savings up to 71\% against H.264 codec and 44\% against H.265 codec in variety of conditions including lighting changes, and rain. RFDVC exhibits greater resilience to transmission errors, outperforming its standard counterpart in non-rainy conditions with up to a +0.26 SSIM improvement (H.264 RFDVC vs. H.264 VC) and maintaining a robust +0.18 improvement in rainy conditions, with both results at a BLER of 0.25. This resilience is further highlighted by the +0.29 SSIM performance gap observed between the H.264 and H.265 RFDVC variants under the non-rainy scenario.\par

Future research will prioritize the processing of downstream tasks in advance within the vehicular metaverse, alongside adaptive updates to RFs as required. This strategy aims to minimize latency and improve virtual environment accuracy, thereby optimizing real-time performance for sophisticated vehicular applications in complex and data-rich scenarios.

\section*{Acknowledgments}
This work has been partially funded by the US National Science Foundation under grant CCF 2140154. This work was also supported by the Ministry of Education, Science, Research and Sport of the Slovak Republic, and the Slovak Academy of Sciences under Grant VEGA 1/0685/23 and by the Slovak Research and Development Agency under
Grant APVV SK-CZ-RD-21-0028 and APVV-23-0512, and by Research and Innovation Authority VAIA under Grant 09I03-03-V04-00395. We would also like to acknowledge that Figs.~\ref{fig:cav_mec}, \ref{fig:video_compression_rf_av} and \ref{fig:fastsam_yolo} were made using licensed vector graphics available at Freepik.com.

\bibliography{bibliography.bib}{}
\bibliographystyle{IEEEtran}


\vspace{11pt}

\begin{IEEEbiography}[{\includegraphics[width=1in,height=1.25in,clip,keepaspectratio]{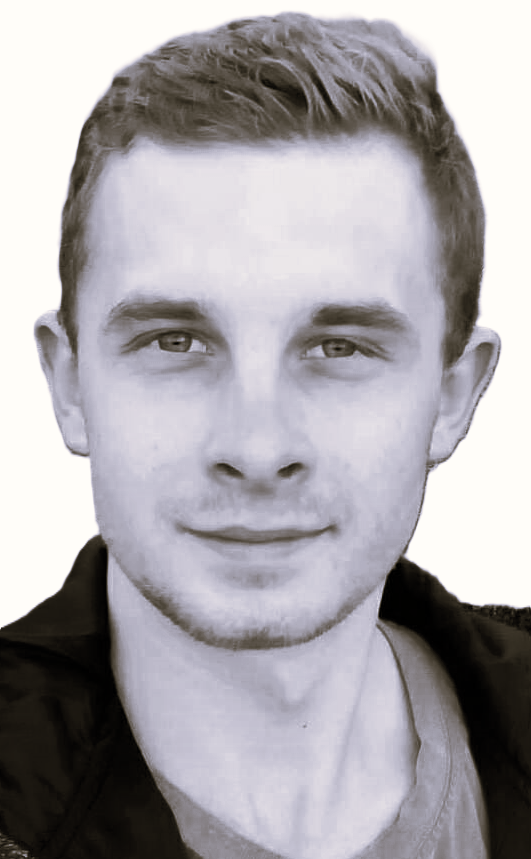}}]{ Matúš Dopiriak} 
 received his BSc. and MSc. degrees in informatics from the Technical University of Ko\v{s}ice, Slovakia. He completed the studies at the Department of Computers and Informatics in the years 2020 and 2022, respectively. His diploma thesis centered on the implementation of CNN-based architectures to monitor traffic from the perspective of an UAV. Currently, he is a doctoral candidate under the supervision of Prof. Ing. Juraj Gazda, PhD. His present research endeavors revolve around the integration of neural architectures, with a specific emphasis on the application of neural radiance fields, in robotics and autonomous mobility of vehicles, particularly in the context of edge computing.
\end{IEEEbiography}

\vspace{11pt} 
\begin{IEEEbiography}[{\includegraphics[width=1in,height=1.25in,clip,keepaspectratio]{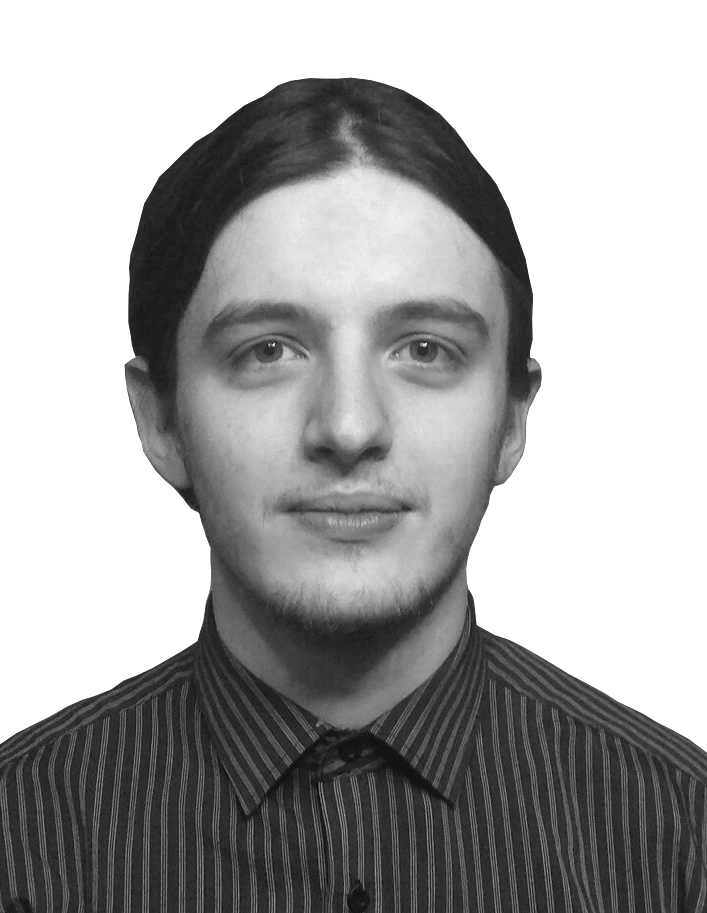}}]{Eugen \v{S}lapak} is an assistant professor at the Technical University of Ko\v{s}ice, Slovakia. His PhD thesis focused on 5G HetNet physical topology design using advanced machine learning algorithms. He was a guest researcher at King`s College London under supervision of Prof. Mischa Dohler. Currently, he works with the research team at the university's Intelligent Information Systems Laboratory (http://iislab.kpi.fei.tuke.sk/), and his research interests include radio access network simulation, computer vision, metaheuristic optimization and machine learning. He serves as the regular reviewer in several recognized IEEE Transactions journals.
\end{IEEEbiography}
\vspace{11pt} 
\begin{IEEEbiography}[{\includegraphics[width=1in,height=1.25in,clip,keepaspectratio]{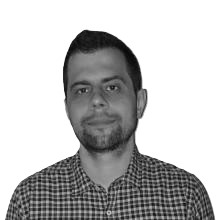}}]{Juraj Gazda} is currently the Vice-Rector for Innovation and Technology Transfer at the Technical University of Košice (TUKE), Slovakia, and a full professor with the Faculty of Electrical Engineering and Informatics at the same university. He has been a guest researcher at Ramon Llull University, Barcelona, and the Technical University of Hamburg-Harburg. He has been involved in the development for Nokia Siemens Networks (NSN) and Ericsson. In 2017, he was recognized as the Best Young Scientist at TUKE. Currently, he serves as the editor of \emph{KSII Transactions on Internet and Information Systems} and as a guest editor of \emph{Wireless Communications and Mobile Computing (Wiley)}. His research interests include techno-economic aspects of 5G/6G networks, computer vision, and artificial intelligence.
\end{IEEEbiography}
\vspace{11pt} 
\begin{IEEEbiography}[{\includegraphics[width=1in,height=1.25in,clip,keepaspectratio]{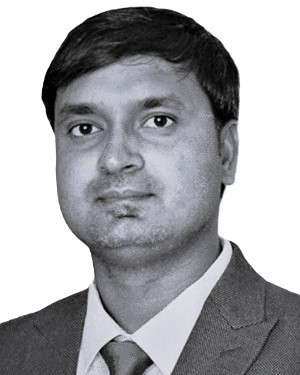}}]{Devendra Singh Gurjar} received the B.Tech. degree in electronics and
communications engineering from Uttar Pradesh
Technical University, Lucknow, India, in 2011, the
M.Tech. degree in wireless communications and
computing from the Indian Institute of Information
Technology Allahabad, India, in 2013. He received
the Ph.D. degree in electrical engineering from
the Indian Institute of Technology Indore, India, in
2017. He was with the department of electrical and
computer engineering, University of Saskatchewan,
Canada, as a Postdoctoral Research Fellow. Currently, he is working as an
Assistant Professor in the department of electronics and communication
engineering, National Institute of Technology Silchar, Assam, India.
He is recipient of Alain Bensoussan Fellowship-2019 from European
Research Consortium for Informatics and Mathematics (ERCIM). With this
fellowship, he worked with the Department of Information Security and
Communication Technology, NTNU Norway. He has numerous publications
in peer-reviewed journals and conferences. His research interests include
MIMO communication systems, cooperative relaying, device-to-device
communications, smart grid communications, physical layer security, and
simultaneous wireless information and power transfer. He is a member of the
IEEE Communications Society and the IEEE Vehicular Technology Society. 
\end{IEEEbiography}
\vspace{11pt} 
\begin{IEEEbiography}[{\includegraphics[width=1in,height=1.25in,clip,keepaspectratio]{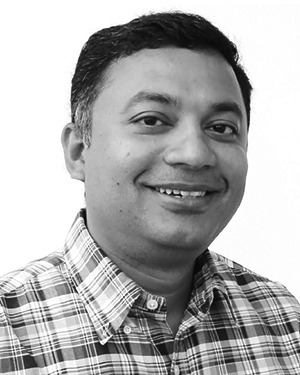}}]
{Mohammad Abdullah Al Faruque} (Senior
Member, IEEE) received the Ph.D. degree in
computer science from Karlsruhe Institute of
Technology, Karlsruhe, Germany, in 2009.
He is currently with the University of California, Irvine, CA, USA, as a Full Professor
and directs the Embedded and Cyber-Physical
Systems Lab and the Samueli School of Engineering Autonomous Systems Initiatives. His
research focuses on the system-level design of
embedded and cyber-physical systems (CPS)
with a special interest in low-power design, CPS security, and data-driven CPS design.
Prof. Al Faruque has received four Best Paper Awards (ACSAC-2022,
DATE-2016, DAC-2015, and ICCAD-2009). He received the IEEE Technical Committee on Cyber-Physical Systems Early-Career Award and
the IEEE CEDA Ernest S. Kuh Early Career Award. He was awarded
the Thomas Alva Edison Patent Award for one of his inventions. He is an
ACM Senior Member. He is also the IEEE CEDA Distinguished Lecturer.
\end{IEEEbiography}
\vspace{11pt} 
\begin{IEEEbiography}[{\includegraphics[width=1in,height=1.25in,clip,keepaspectratio]{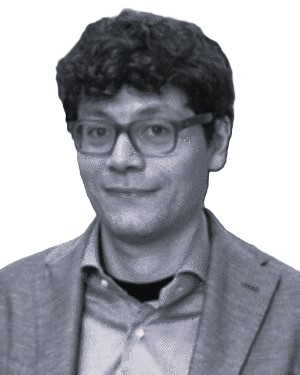}}]
{Marco Levorato} (Senior Member, IEEE) received
 the Ph.D. degree in electrical engineering from the
 University of Padova, Padua, Italy, in 2009. In August
 2013, he joined the Computer Science Department,
 University of California Irvine, Irvine, CA, USA.
 Between 2010 and 2012, he was a Postdoctoral Re
searcher with a joint affiliation with Stanford and
 the University of Southern California, working with
 Prof. Andrea Goldsmith and Prof. Urbashi Mitra.
 From January to August 2013, he was an Access
 Postdoctoral affiliate with the Access center, Royal
 Institute of Technology, Stockholm. He has coauthored more than 100 technical
 articles on these topics, including the paper that was the recipient of the Best
 Paper Award at IEEE GLOBECOM (2012). His research interests include
 next-generation wireless networks, autonomous systems, Internet of Things,
 e-health and stochastic control. In 2016, he was the recipient of the UC Hellman
 Foundation Award for his research on Smart City IoT infrastructures. He is a
 Member of the ACM and IEEE ComSoc society.
\end{IEEEbiography}

\vfill

\end{document}